\documentclass[10pt,aps,amsfonts,amssymb,longbibliography,twocolumn]{revtex4-1}
\usepackage{mathrsfs} 
\usepackage{graphicx,color}
\usepackage{hyperref}
\usepackage{bm}
\begin{document}

\title{Replica symmetry breaking in the RKKY skyrmion crystal system}

\author{Kota Mitsumoto}
\author{Hikaru Kawamura}

\affiliation{Molecular Photoscience Research Center, Kobe University, Kobe 657-8501, Japan}

\begin{abstract}
We study the RKKY Heisenberg model on a three-dimensional stacked-triangular lattice under magnetic fields by extensive Monte Carlo simulations to get insight into the chiral-degenerate symmetric skyrmion crystal (SkX) in centrosymmetric metallic magnets. The triple-$q$ SkX state and the double-$q$ states are realized, together with the single-$q$ state. We find an unexpected phenomenon of the replica-symmetry breaking (RSB) well-known in glassy systems, although the Hamiltonian and the ordered state are entirely regular. In the RSB SkX phase, the triple-$q$ SkX state macroscopically coexists with the single-$q$ state, in spite of the fact that these ordered states cannot be transformed via any Hamiltonian-symmetry operation. In the thermodynamic limit, the free energies of these states are degenerate whereas the free-energy barrier between the states diverges, breaking the ergodicity. A similar RSB is observed also in the RSB double-$q$ phase where the double-$q$ state macroscopically coexists with the single-$q$ state. Experimental implications are discussed.
\end{abstract}

\maketitle

\section{introduction}

 Frustrated magnets have been extensively studied because of their unconventional behaviors such as spin liquids \cite{anderson1973resonating, balents2010spin, zhou2017quantum, kawamura2019nature}, spin glasses  \cite{mezard1987spin, mydosh1993spin, kawamura2015spin} and chiral ordered states \cite{miyashita1984nature, okubo2012multiple}. Frustration often suppresses ordinary magnetic long-range orders, and instead induces various intriguing magnetic states. Much attention has recently been paid to emergent nano-scale spin objects, e.g. topological spin textures including the vortex \cite{kamiya2014magnetic, ozawa2016vortex, hayami2018neel}, the $Z_2$ vortex \cite{kawamura1984phase,kawamura2010z}, the skyrmion \cite{okubo2012multiple, ozawa2017zero, hayami2017effective, lin2018face, wang2020skyrmion} and the hedgehog \cite{okumura2020magnetic, aoyama2021hedgehog}. Such interest has also been amplified by possible applications to spin-electronic devices \cite{nagaosa2013topological, fujishiro2018large}.

 The skyrmion is a swirling noncoplanar texture whose constituent spin directions wrap a sphere in spin space, characterized by the integer topological charge in units of the solid angle $4\pi$. The sign of the topological charge represents the swirling direction of the skyrmion, or the sign of the scalar spin chirality representing the handedness of the associated noncoplanar spin structure. The skyrmion has so far been observed mostly in the Dzyaloshinskii-Moriya (DM) interaction systems such as MnSi, FeCoSi or FeGe in the form of the skyrmion crystal (SkX), the periodic array of skyrmions \cite{muhlbauer2009skyrmion, yu2010real, yu2011near}. These systems exhibit the DM-interaction-induced spiral structure in zero magnetic field, and exhibit the SkX under finite magnetic fields. Especially interesting aspect of the SkX state might be its nontrivial electromagnetic response. Nonzero total scalar chirality of the SkX leads to the anomalous Hall conductivity \cite{neubauer2009topological}, the so-called topological Hall effect. Note that, since the DM interaction breaks the chiral degeneracy of the Hamiltonian energetically discriminating the right-handed and left-handed spirals, the SkX of the DM system possesses a definite sign of the topological charge.

 Recent studies have revealed that the SkX phase can also be realized even in the centrosymmetric magnets without the DM interaction, which is induced by the frustrated exchange interaction, e.g. the classical $J_1$-$J_3$ (or $J_1$-$J_2$) Heisenberg model on the triangular lattice with the competing nearest-neighbor ($J_1$) and further-neighbor interactions ($J_2$, $J_3$) \cite{okubo2012multiple}. The model shows the spiral single-$q$ structure in zero magnetic field, and the multiple-$q$ structures under magnetic fields, including the double-$q$ phase and the triple-$q$ phase. The latter corresponds to the SkX phase. Interestingly, in sharp contrast to the DM-induced skyrmions, both skyrmions and anti-skyrmions characterized by the mutually opposite signs of the topological charge (scalar spin chirality) are possible due to the $Z_2$ chiral degeneracy of the Hamiltonian, either the SkX state or the anti-SkX state being selected via the spontaneous symmetry breaking. Such $Z_2$ chiral degeneracy inherent to the frustration-induced SkX also gives rise to a peculiar phase, called the $Z$ phase, which is the random domain state consisting of both SkX and anti-SkX. A similar chiral degenerate SkX state has been reported also in the frustrated classical Heisenberg model on the three-dimensional (3D) stacked-triangular lattice \cite{osamura}. The effect of the easy-axis-type magnetic anisotropy in the 3D models was also extensively studied \cite{lin2018face}.

 Recently, some candidate materials of the frustration-induced SkX were reported, such as Gd$_2$PdSi$_3$ \cite{saha1999magnetic, kurumaji2019skyrmion} and EuCuSb \cite{takahashi2020competing}. They are mostly metallic compounds with more or less 3D interactions. Their magnetism is born by localized magnetic moments which are coupled via the Ruderman-Kittel-Kasuya-Yosida (RKKY) interaction mediated by itinerant electrons. In contrast to the short-range interaction inherent to insulating magnets, the RKKY interaction is the long-range interaction falling off as $1/r^3$ and oscillating in sign with the distance $r$ \cite{ruderman1954indirect, kasuya1956theory, yosida1957magnetic}. While the oscillating nature of the RKKY interaction certainly provides frustration, it remains open whether the chiral-degenerate symmetric SkX state is really stabilized under magnetic fields in the RKKY system, and if so, how its nature could differ from the one stabilized in the short-range system. 

 In the present paper, in order to address this question, especially to elucidate the effects of the long-range nature and the 3D character of the interaction inherent to  metallic experimental systems, we investigate by means of extensive Monte Carlo (MC) simulations the properties of the classical Heisenberg model on the 3D stacked-triangular lattice interacting via the long-range RKKY interaction.

 Recently, possible phases in metallic systems were theoretically investigated by numerical simulations \cite{wang2020skyrmion, ozawa2017zero, hayami2017effective}. In Ref.\cite{wang2020skyrmion}, the 2D Heisenberg model interacting with the RKKY interaction mediated by 2D electron gas was investigated by means of the variational calculation, and various phases including the SkX state were reported in the ground state. In Refs.\cite{ozawa2017zero, hayami2017effective}, the possible phases were investigated by numerical simulation on the Kondo-lattice model \cite{ozawa2017zero} and by the analysis of the effective mean-field-type classical spin model in $q$-space \cite{hayami2017effective}. The SkX was reported to be stabilized by the contribution of the higher-order four-body (biquadratic) spin-spin interaction with the positive coupling constant \cite{hayami2017effective}. Such a situation might be expected when the exchange interaction between itinerant electrons and localized moments is comparable to the Fermi energy. Our present modeling presumes the more weak-coupling situation where the exchange interaction is smaller enough than the Fermi energy, resulting in the dominant quadratic spin-spin interaction, {\it i.e.\/}, the standard RKKY interaction.
 
 Via extensive MC simulations, we find that the chiral-degenerate symmetric triple-$q$ SkX state is stabilized in the 3D RKKY system for a certain parameter range of the RKKY interaction. The phase diagram turns out to be not much different from the one of the 2D short-range systems reported earlier \cite{okubo2012multiple}. However, the nature of the SkX turns out to be quite different from that of the 2D short-range system. Namely, we find an unexpected phenomenon of the so-called replica-symmetry breaking (RSB), in sharp contrast to the case of the 2D short-range model.

 The notion of the RSB was first introduced to solve the mean-field spin-glass model in infinite dimensions by using the replica method \cite{mezard1987spin, parisi1979infinite}, which is a theoretical tool to treat the quenched disorder of the Hamiltonian. Though the original interpretation of the RSB was rather technical, the physical interpretation of the RSB now widely accepted might be that multiply-degenerate free-energy minima which are {\it unrelated\/} via any global symmetry operation of the underlying Hamiltonian emerge in the phase space, quite distinct from the spontaneous symmetry breaking in the usual phase transition where degenerate free-energy minima related via the global symmetry operation of the underlying Hamiltonian emerge accompanied with the divergent free-energy barrier. Thus, in case of RSB, many different macroscopic ordered states unrelated by the Hamiltonian symmetry are realized in the thermodynamic limit with the divergent free-energy barriers among them, leading to the broken ergodicity.
 
 More recently, the replica method has been extended and successfully applied to regular systems without quenched disorder, e.g. molecular or structural glasses \cite{charbonneau2014fractal} and disorder-free frustrated magnets \cite{yoshino2018disorder}. Even in such regular systems, the RSB could also occur in the large dimensional limit, and some simulation results supported the prediction of the mean-field theory even in 3D \cite{berthier2016growing, jin2017exploring, mitsumoto2020spin}. Thus, although the RSB is possible even for regular systems without quenched disorder, the states themselves are still glassy, i.e., spatially random and infinitely degenerate. 

 In sharp contrast, the RSB SkX phase we have observed here consists of regular states with spatial periodicity, realized under the completely regular Hamiltonian without quenched disorder: It is a thermodynamic phase in full equilibrium where the triple-$q$ SkX state macroscopically coexists with the single-$q$ spiral state in the sense of RSB. Since the triple-$q$ SkX state cannot be transformed into the single-$q$ spiral state, nor vice versa, by any global symmetry operation of the Hamiltonian, the associated symmetry breaking should be the RSB.

 In addition to such an RSB triple-$q$ SkX phase, we find another RSB phase of slightly different nature, i.e., the RSB double-$q$ state where the double-$q$ state and the single-$q$ state macroscopically coexist together with the spectrum of intermediate states connecting the double-$q$ and the single-$q$ states.

 The rest of the paper is organized as follows. In Sect. \ref{model}, we introduce our model and explain the MC simulation method employed. Sect. \ref{result} is the main part of the paper, where we present the results of our MC simulations including the phase diagram of the model in the temperature versus magnetic-field plane and the behaviors of various physical quantities. Particular attention is paid to the nature of the two types of RSB phases. The nature of the RSB SkX phase is analyzed in detail in subsection A, while that of the RSB double-$q$ phase is analyzed in subsection C. Subsection B is devoted to the dynamical simulation performed to further examine the nature of the RSB SkX phase. Summary and discussion of the results are given in Sect. \ref{summary}. Some of the details of the Ewald sum method is given in Appendix \ref{sec_ewald_app}, additional simulation data not presented in the main text are given in Appendix \ref{sec_mag_temp_app}, typical spin and chirality configurations of the RSB double-$q$ state are given in Appendix \ref{sec_double_spin_chiral}, and the mean-field analysis performed to better understand the RSB double-$q$ phase are given in Appendix \ref{sec_mean_app}.

\section{The model and the method}
\label{model}

\begin{figure}[t]
\includegraphics[clip,width=85mm]{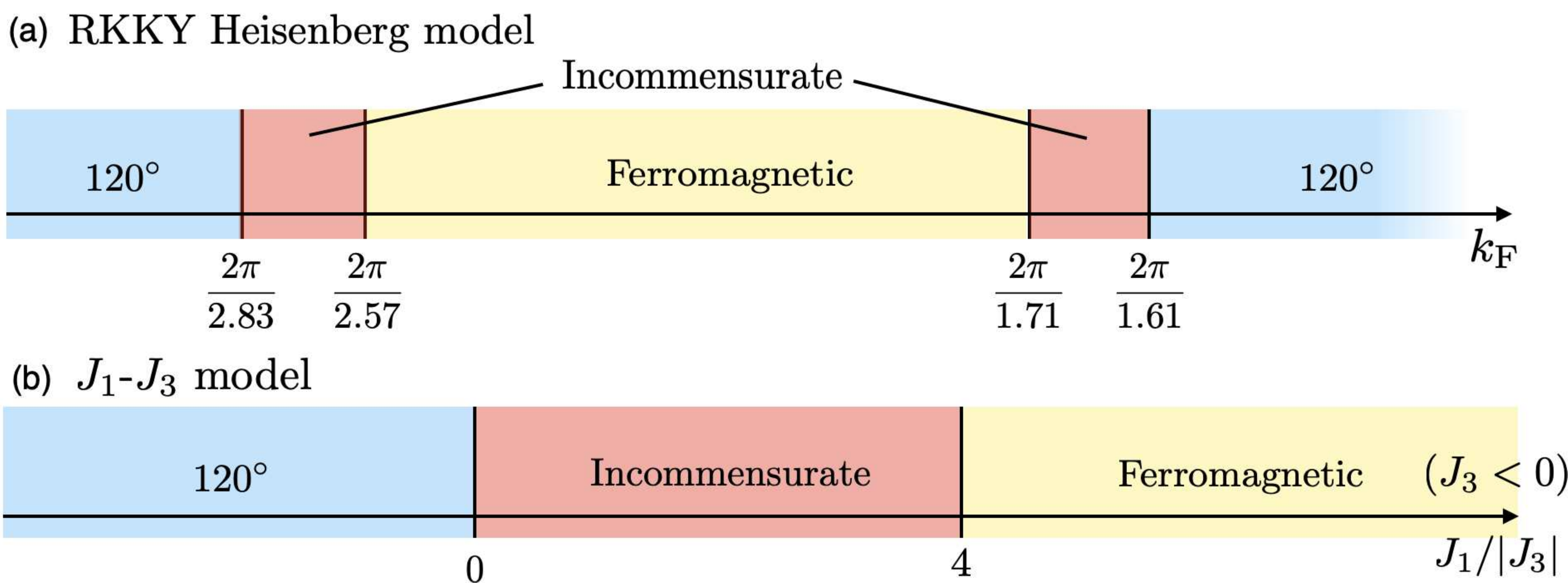}
\caption{The ground-state in-plane phase structures of the frustrated Heisenberg models: (a) The $k_{\rm F}$ dependence for the RKKY classical Heisenberg model on the stacked-triangular lattice with $c = 1.45$, where ``$120^\circ$'' means the commensurate $120^\circ$ spin state with the three-sublattice periodicity, while ``incommensurate'' means the incommensurate spiral states on the triangular plane.  (b) The $J_1/|J_3|$ dependence for the $J_1$-$J_3$ ($J_1>0$, $J_3<0$) classical Heisenberg model on the 2D triangular lattice.
}
\label{inter}
\end{figure}

 We consider the classical Heisenberg model on the 3D stacked triangular lattice interacting with the long-range RKKY interaction. The RKKY interaction is the primary interaction between localized spins in metals originating from the $s$-$d$ coupling between itinerant electrons and localized magnetic moments \cite{ruderman1954indirect, kasuya1956theory, yosida1957magnetic}. Assuming the spherical Fermi surface for itinerant electrons and the $s$-$d$ coupling being sufficiently weak compared with the Fermi energy, we derive the conventional long-range RKKY interaction between two spins with the distance $r$ in the second order perturbation with respect to the $s$-$d$ coupling as,
\begin{equation}
J_{\rm RKKY}(r) \sim  \left( \frac{\cos(2k_{\rm F} r)}{(2k_{\rm F} r)^3} - \frac{\sin(2k_{\rm F} r)}{(2k_{\rm F} r)^4} \right)  ,
\end{equation}
where $k_{\rm F}$ is the Fermi wavevector.

 The RKKY Heisenberg Hamiltonian employed in our MC simulations are given by
\begin{eqnarray}
H&=&-\sum_{i,j}J_{ij}\bm{S}_i \cdot \bm{S}_j - h \sum_{i=1}^N S_i^{z} , \nonumber \\
J_{ij} &=& -J_0 a^3 \left( \frac{\cos(2k_{\rm F} r_{ij})}{r_{ij}^3} - \frac{\sin(2k_{\rm F} r_{ij})}{2k_{\rm F} r_{ij}^4} \right) , 
\label{hamiltonian}
\end{eqnarray}
where $a$ is the lattice constant in the basal $a$-$b$ plane, $r_{ij} = |\bm{r}_i - \bm{r}_j|$ is the distance between the sites $i$ and $j$, and $J_0$ is the energy scale of the interaction,  $k_{\rm F}$ being given is in units of $a^{-1}$. The sum is taken over all spin pairs on the lattice, while the position of the lattice site $i$ is given by,
\begin{eqnarray}
\bm{r}_i &=& n_a \bm{a} + n_b \bm{b} + n_c \bm{c}, \label{position} \\
\bm{a} &=& (1, 0, 0), ~\bm{b} = (1/2,\sqrt{3}/2 , 0), ~\bm{c} = (0,0,c) ,
\end{eqnarray}
where $c$ is the lattice-constant ratio of the interplane to the intraplane ones, $n_a,n_b=0,1,2,...,L-1,~n_c = 0,1,2,...,L_z-1$ are integers, and the total number of lattice sites of the lattice is given by $N=L\times L \times L_z$.

 In addition to the dimensionless temperature $\tilde{T} = k_{\rm B}T/(a^3J_0)$ and the dimensionless magnetic field $\tilde{h} = h/(a^3J_0)$ (hereafter we call $\tilde T$ and $\tilde h$ simply as $T$ and $h$), the model has two characteristic dimensionless parameters, i.e., the Fermi wavenumber $k_{\rm F}$ and the lattice-constant ratio $c$. Then, the ground state in zero field is determined by the set of $k_{\rm F}$ and $c$.

 For the present classical Heisenberg model, the ground-state properties in zero field can be obtained by the simple Fourier transform, and the resulting ground-state spin configurations in the triangular plane are given in Fig. \ref{inter}, as compared with those of the short-range $J_1$-$J_3$ model on the 2D triangular lattice. In addition to the ferromagnetic state and the commensurate $120^\circ$ state, incommensurate spiral structures are stabilized for a certain range of parameters as in the case of the  $J_1$-$J_3$ model. 

 In this study, we concentrate on the parameter range corresponding to the incommensurate structure, and the simulation is performed for the fixed parameters $k_{\rm F} = 2\pi/2.77$ and $c = 1.45$ lying in this incommensurate region. For this case, the ground state in zero field takes the incommensurate spiral structure in the $a$-$b$ plane and the ferromagnetic uniform structure along the $c$ axis.

 The lattice sizes studied in our MC simulations are $L=20, 24, 30$ and $L_z = L/2$, periodic boundary conditions applied in all directions. To take account of the long-range nature of the RKKY interaction beyond the finite system size $L$, especially to avoid the uncertainty associated with a finite-range cutoff of the long-range interaction, we employ the Ewald sum method which is a general method to treat the long-range interaction in numerical simulations \cite{ewald_p_p_1921_1424363, hansen1973statistical, fuchizaki1994towards, ikeda2008ordering}. In the method, we sum over the interactions to infinite range by periodically placing copy images of the finite system in all directions. Details of the derivation of the Ewald potential for the RKKY interaction on a stacked triangular lattice is given in Appendix \ref{sec_ewald_app}.

 To equilibrate the system, we make a combined use of the Metropolis method, the over-relaxation method and the replica-exchange method. Our unit Monte Carlo step (MCS) consists of one Metropolis sweeps and $L/2$ times over-relaxation sweeps. The replica-exchange trial is performed every five MCS. We take $1.0 \times10^5$ MCS each for equilibration and for taking thermal averages. The errors of the physical quantities were evaluated over three independent runs. In some special occasions, we turn off the replica-exchange process, or both the replica-exchange and the over-relaxation processes. One such case corresponds to the measurement of the spin structure factor, and the other corresponds to the dynamical simulations in equilibrium, as will be referred to when employed.

\section{The results}
\label{result}

\begin{figure}[t]
\includegraphics[clip,width=87mm]{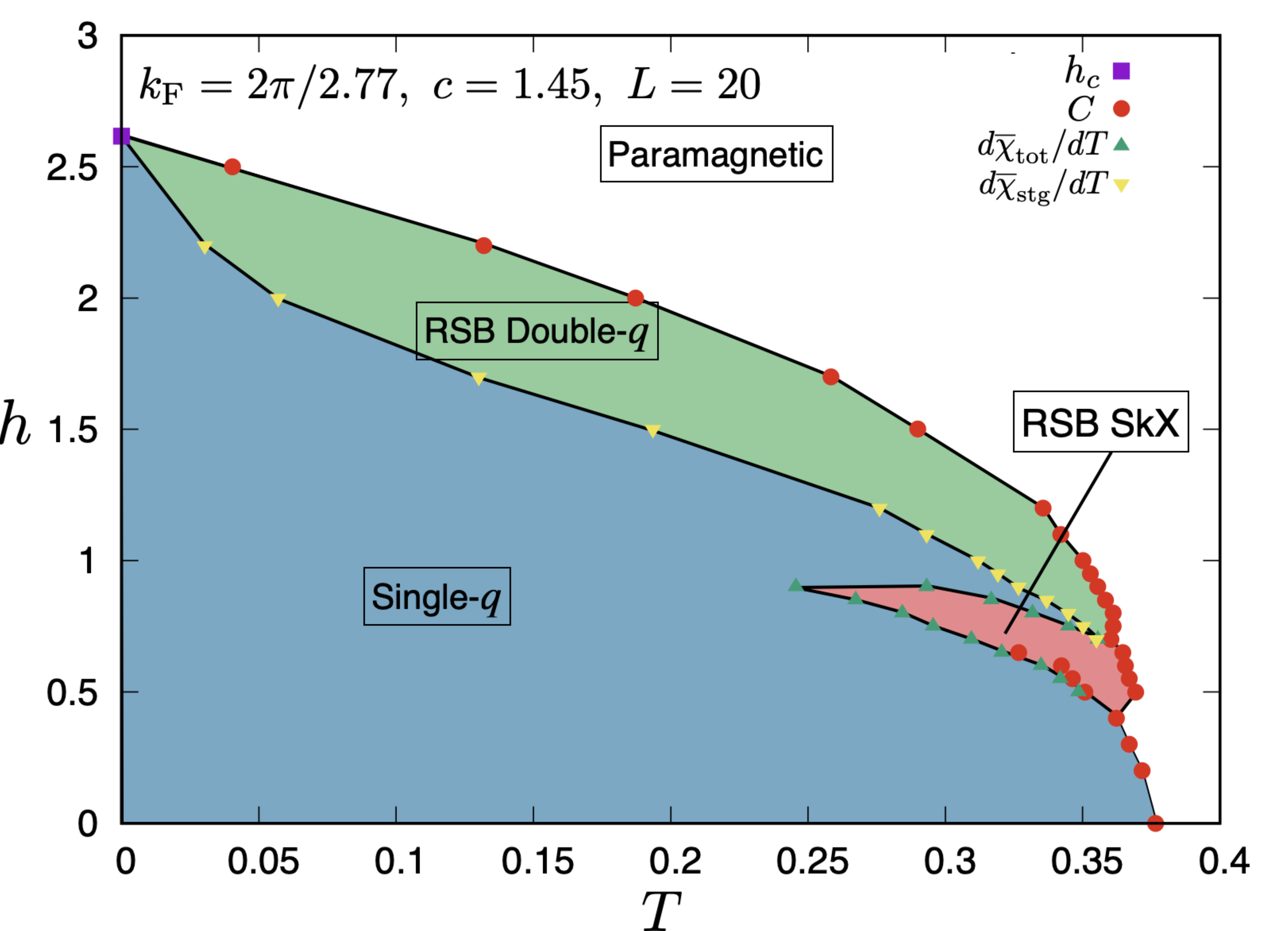}
\caption{The phase diagram in the temperature ($T$) versus magnetic-field ($h$) plane of the RKKY classical Heisenberg model on the stacked-triangular lattice with $k_{\rm F} = 2\pi/2.77$ and $c=1.45$. The phase boundaries are determined from the specific-heat peak ($C$), the $T$-derivative of the mean total scalar chirality ($d\overline{\chi}_{\rm tot}/dT$) and the $T$-derivative of the mean staggered scalar chirality ($d\overline{\chi}_{\rm stg}/dT$) of the $L=20$ data. $T=0$ transition point ($h_c$) is determined from the analytic result of the ground-state energy.
}
\label{phase}
\end{figure}

\begin{figure}
\includegraphics[clip,width=85mm]{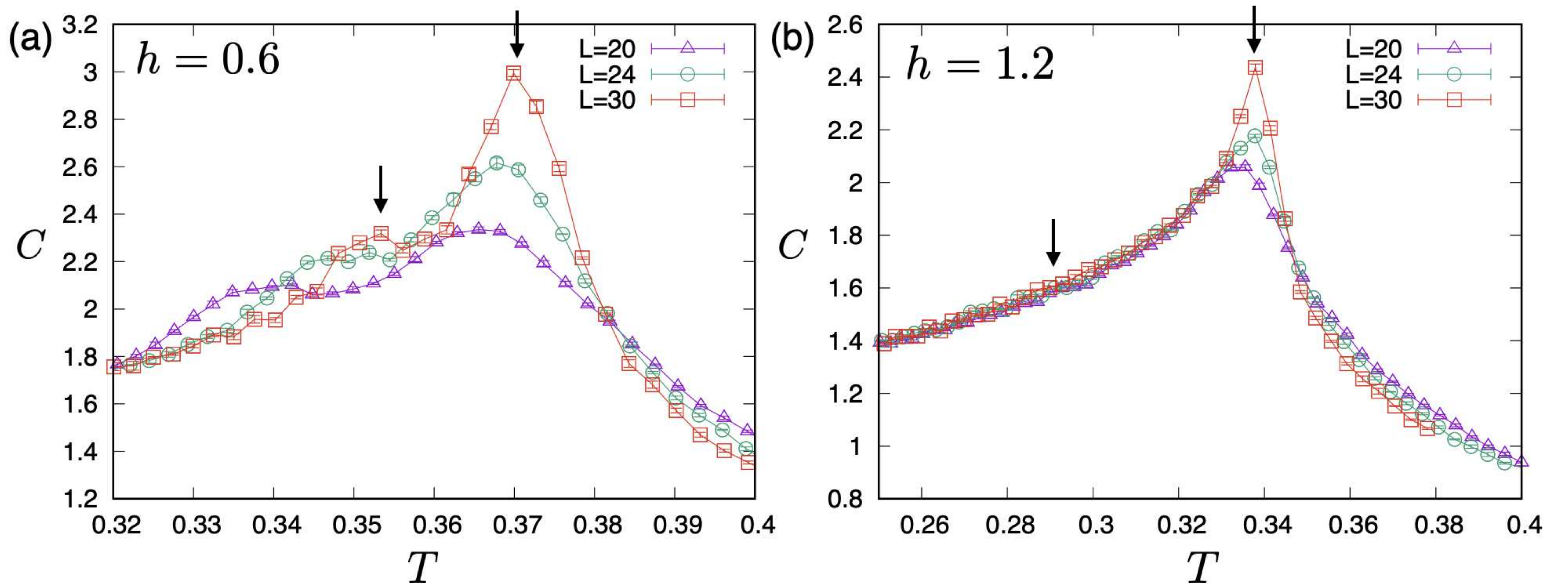}
\caption{The temperature dependence of the specific heat under applied fields of (a) $h=0.6$, and (b) $h=1.2$, each crossing the RSB SkX phase and the RSB double-$q$ phase.
}
\label{heatcap}
\end{figure}

 In this section, we present our simulation results. Fig. \ref{phase} shows the phase diagram of the model  in the temperature ($T$) versus magnetic-field ($h$) plane, for the parameters $k_{\rm F} = 2\pi/2.77,~c=1.45$ where the incommensurate helical spin structure is realized in the zero-field ground state. In addition to the single-$q$ phase, which occupies a wide region of the phase diagram, the ``double-$q$'' and ``the triple-$q$ SkX'' phases are also stabilized under fields. The overall geometry of the phase diagram looks rather similar to that of the 2D short-range model, except that the $Z$ phase stabilized in the 2D short-range model \cite{okubo2012multiple} seems absent in the present model, and that the single-$q$ state extended toward higher temperatures along the high-field phase boundary in the 2D short-range model is absent in the present case. Nevertheless, the characters of each phase, especially those of the ``double-$q$'' phase and the ``triple-$q$ SkX'' phase are quite different from the short-model counterparts, i.e., these multiple-$q$ phases exhibit the RSB, macroscopically coexisting with the single-$q$ state as will be detailed below. Hence, these multiple-$q$ phases are described as ``RSB double-$q$'' and ``RSB SkX'' in the magnetic phase diagram of Fig. \ref{phase}.

 The phase boundary of Fig. \ref{phase} is drawn from the anomaly of several physical quantities including the specific heat by using the $L=20$ data. As an example, we show in Fig. \ref{heatcap} the temperature dependence of the specific heat for several sizes $L$ for the case of the magnetic field $h=0.6$ (Fig. \ref{heatcap} (a)) and $h=1.2$ (Fig. \ref{heatcap} (b)), each crossing the RSB SkX phase or the RSB double-$q$ phase when the temperature is lowered from the high-$T$ paramagnetic phase toward the low-$T$ single-$q$ phase. While the transition points $T_c$ between the paramagnetic/RSB-SkX phases,  between the paramagnetic/RSB-double-$q$ phases, and between the RSB-SkX/single-$q$ phases are eminent from Figs
. \ref{heatcap} (a) and \ref{heatcap} (b), the transition point between the RSB-double-$q$/single-$q$ phases expected for $h=1.2$ is hardly visible from Fig. \ref{heatcap} (b) However, it can clearly be identified from other quantities as shown below. Additional data of various physical quantities are also given in Appendix \ref{sec_mag_temp_app}.

\subsection{RSB SkX Phase}

 Now, we investigate the properties of each phase in more detail, especially focusing on the RSB character of the RSB triple-$q$ SkX phase and the RSB double-$q$ phase.

 We begin with the RSB SkX phase, the magnetic field being fixed to a typical value of $h=0.6$. The RSB SkX phase is stabilized in the intermediate temperature region $T_{\rm c1} > T > T_{\rm c2}$. While $T_{{\rm c1}}$ is estimated to be $T_{{\rm c1}}\simeq 0.37$ for $h=0.6$, $T_{{\rm c2}}$ is more subject to the finite-size effect but its upper limit can be estimated to be $T_{{\rm c2}}\lesssim 0.358$: See below.

\begin{figure}[t]
\includegraphics[clip,width=80mm]{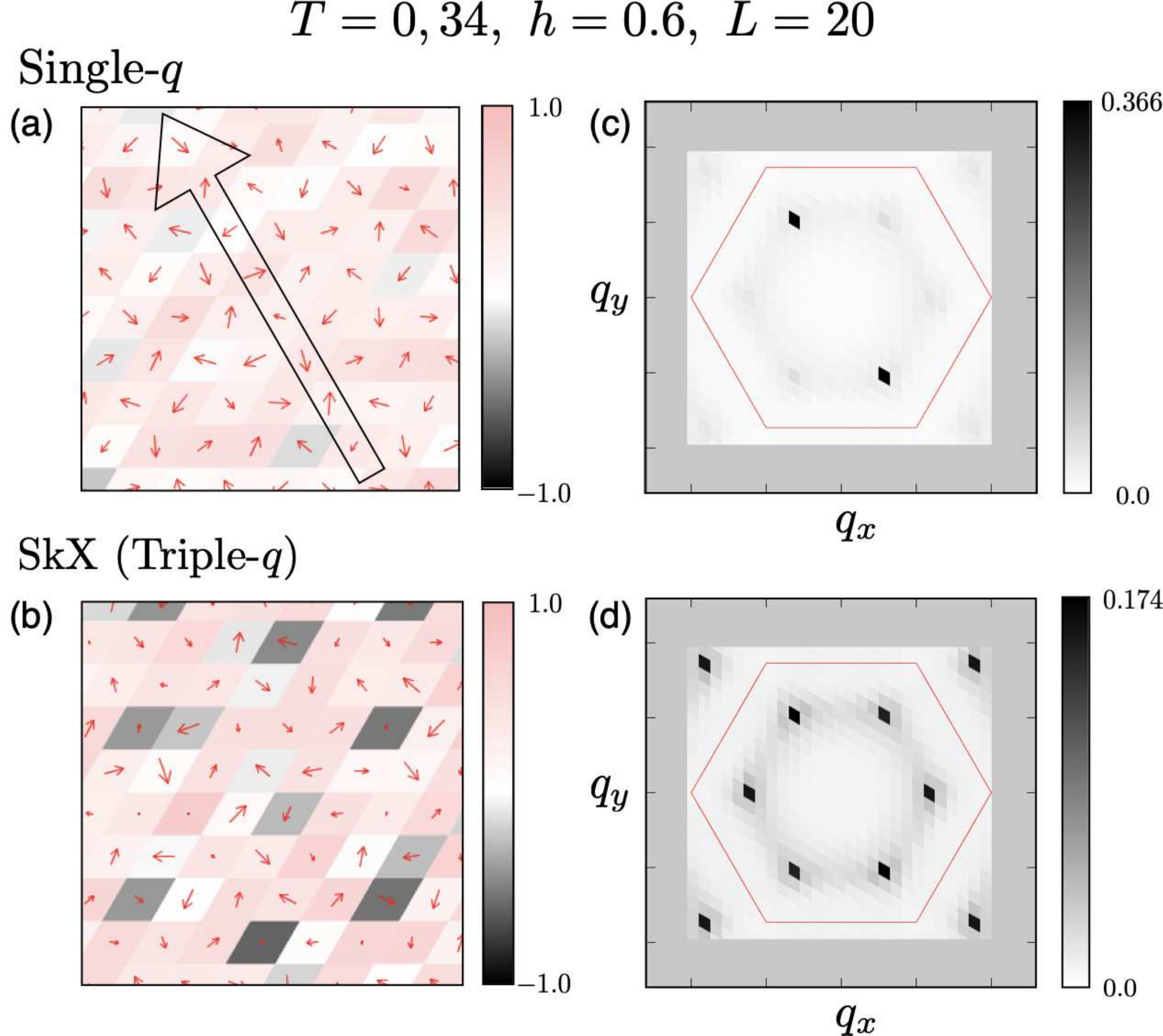}
\caption{Typical spin configurations realized as an equilibrium state in the RSB SkX phase at ($T=0.34, h=0.6$), given in real space [left column] and in $q$-space [right column]. The lattice size is $L=20$. The real-space spin configurations in the triangular plane of (a) the single-$q$ state, and of (b) the triple-$q$ SkX state, each realized as an equilibrium state. To reduce the thermal noise, the short-time averaging over $100$ MCS with only the Metropolis updating is made. The colors of the rhombuses represent the $S_z$-component of spin and the arrows represent the ($S_x,S_y$)-components. Along the orthogonal direction, these spin configurations are stacked ferromagnetically. The perpendicular static spin structure factor $S_\perp (\bm{q})$ in the ($q_x$, $q_y$)-plane with $q_z = 0$ of (c) the single-$q$ state, and of (d) the SkX state. The red hexagon represents the first Brillouin zone. In measuring $S_\perp (\bm{q})$, MC simulations are made with only the Metropolis updating averaged over $10^4$ MCS.
}
\label{snap_skx}
\end{figure}

\begin{figure}[t]
\includegraphics[clip,width=87mm]{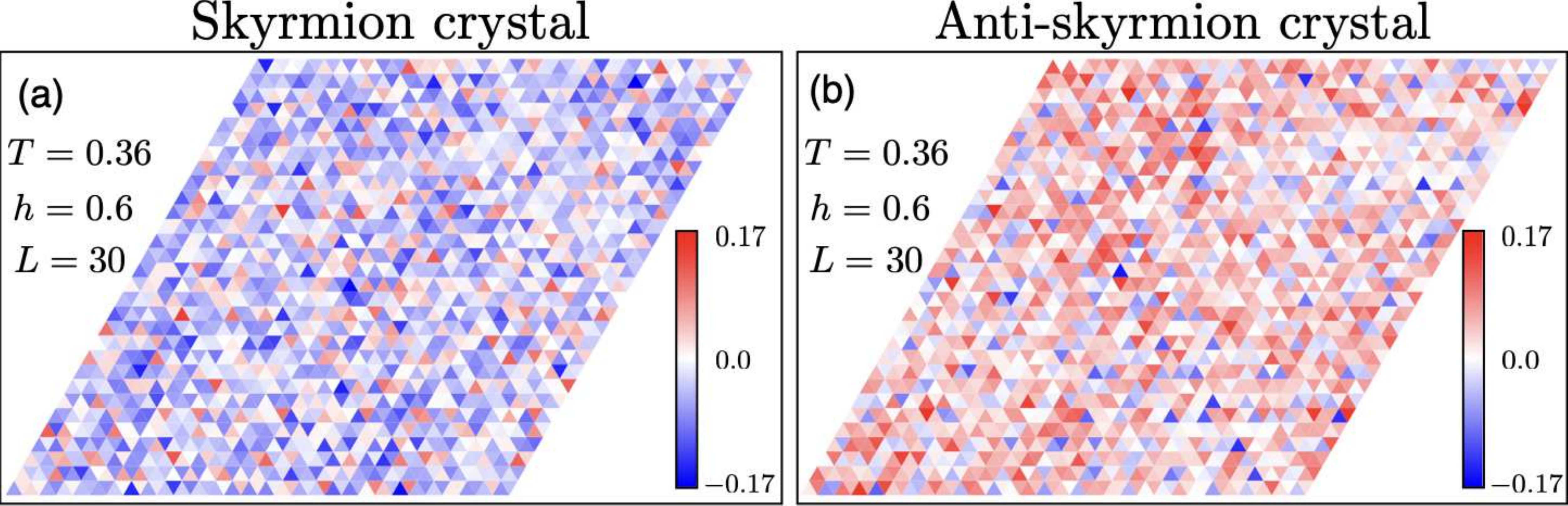}
\caption{Typical spatial distributions of the scalar chirality in (a) the SkX state, and in (b) the anti-SkX state, realized as an equilibrium state in the RSB SkX phase at ($T=0.36, h=0.6$) for the lattice size $L=30$. To reduce the thermal noise, the short-time averaging over 100 MCS with only the Metropolis updating is made. The color represents the sign of the scalar chirality.
}
\label{snapchiral_skx}
\end{figure}

  In the temperature range $T_{{\rm c1}} > T > T_{{\rm c2}}$, in the course of our replica-exchange MC simulations at full equilibrium we observe both the single-$q$ state and the triple-$q$ SkX state, each appearing with finite probability. Note that both states are observed as spatially uniform states, any domain state consisting of domains hardly observed. At ($T=0.34$, $h=0.6$) located in the middle of the RSB triple-$q$ SkX phase, we show in Fig. \ref{snap_skx} the real-space spin configurations (short-time averaged to reduce the thermal noise) and the spin structure factor in the ($q_x$, $q_y$)-plane with $q_z=0$ observed in the course of MC simulation for each state, i.e., the upper row corresponding to the single-$q$ state and the lower row to the the triple-$q$ SkX state. The spin structure factors for the field-perpendicular and field-parallel spin components are defined by
\begin{eqnarray}
S_\perp (\bm{q}) &=& \frac{1}{N} \left\langle \left( \sum_{\mu = x,y} \left| \sum_{i=1}^N S_i^\mu e^{-i \bm{q}\cdot \bm{r}_i}\right|^2 \right)^{1/2} \right\rangle, \label{insta_perp} \\
S_\parallel (\bm{q}) &=& \frac{1}{N} \left\langle \left( \left| \sum_{i=1}^N S_i^z e^{-i \bm{q}\cdot \bm{r}_i} \right|^2 \right)^{1/2} \right\rangle. \label{insta_para}
\end{eqnarray}
As can be seen from the figures, both the single-$q$ state and the triple-$q$ SkX state look quite similar to the corresponding standard single-$q$ and triple-$q$ SkX states observed earlier in the 2D short-range model. The unique feature of the present case is that these states are simultaneously stabilized as a result of RSB at a common ($T,h$) point in the phase diagram, in sharp contrast to the standard case where each state is realized at different ($T,h$) points in the phase diagram in different phases. 

 Aside from the RSB, more standard spontaneous symmetry breaking associated with the Hamiltonian symmetry also takes place in the RSB triple-$q$ state. For example, the triple-$q$ SkX state spontaneously breaks the $Z_2$ mirror symmetry in the spin space, e.g., the symmetry under the operation $(S_x,S_y,S_z)$ to  $(-S_x,S_y,S_z)$, and as a result, both the SkX state and the anti-SkX state characterized by the mutually opposite signs of the scalar chirality are equally possible. This is demonstrated in Fig. \ref{snapchiral_skx} where both the SkX and the anti-SkX states are shown, each realized in the RSB triple-$q$ SkX state. As pointed out in Ref.\cite{okubo2012multiple}, the realization of both the SkX and the anti-SkX states with mutually opposite topological charge, and the resulting mutually opposite electromagnetic responses, are interesting features of the frustration-induced chiral-degenerate SkX state different from the DM-induced SkX state. The present RKKY system certainly shares this interesting feature even in the occurrence of the RSB.

\begin{figure*}
\centering
\includegraphics[clip,width=175mm]{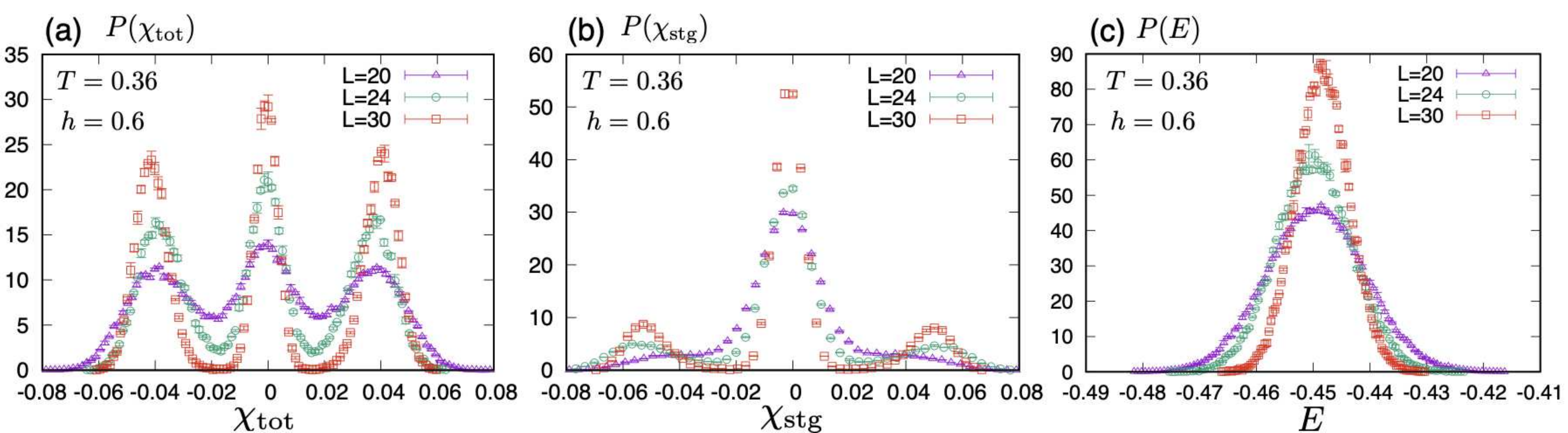}
\caption{Equilibrium distribution functions for various sizes $L$ in the RSB SkX phase at ($T=0.36,~h=0.6$) of (a) the total scalar chirality, (b) the staggered scalar chirality, and (c) the energy per spin.
}
\label{dist_skx}
\end{figure*}

\begin{figure}[b]
\centering
\includegraphics[clip,width=85mm]{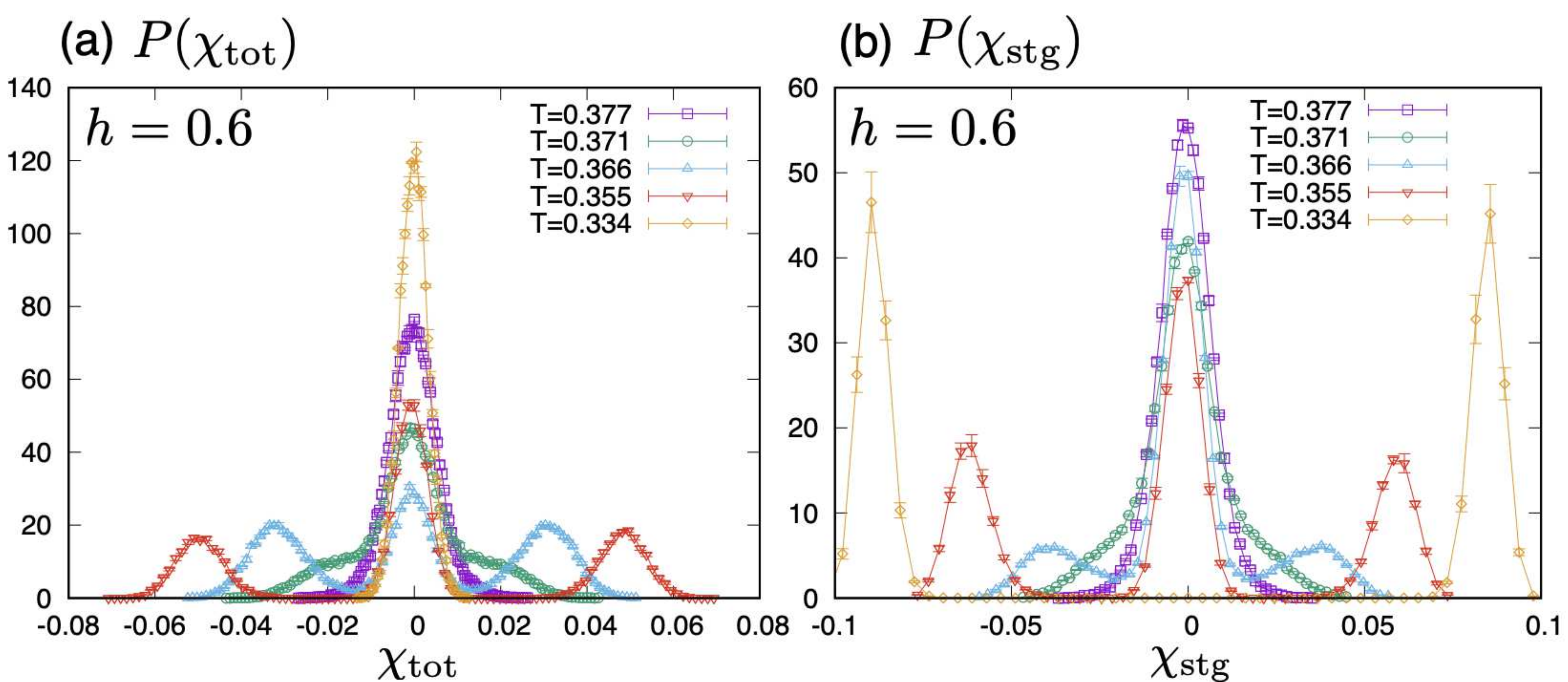}
\caption{The temperature dependence of the chirality distribution functions at a field $h=0.6$ for the size $L=30$: (a) The total scalar chirality, and (b) the staggered scalar chirality.
}
\label{dist_skx2}
\end{figure}

\begin{figure}[b]
\includegraphics[clip,width=85mm]{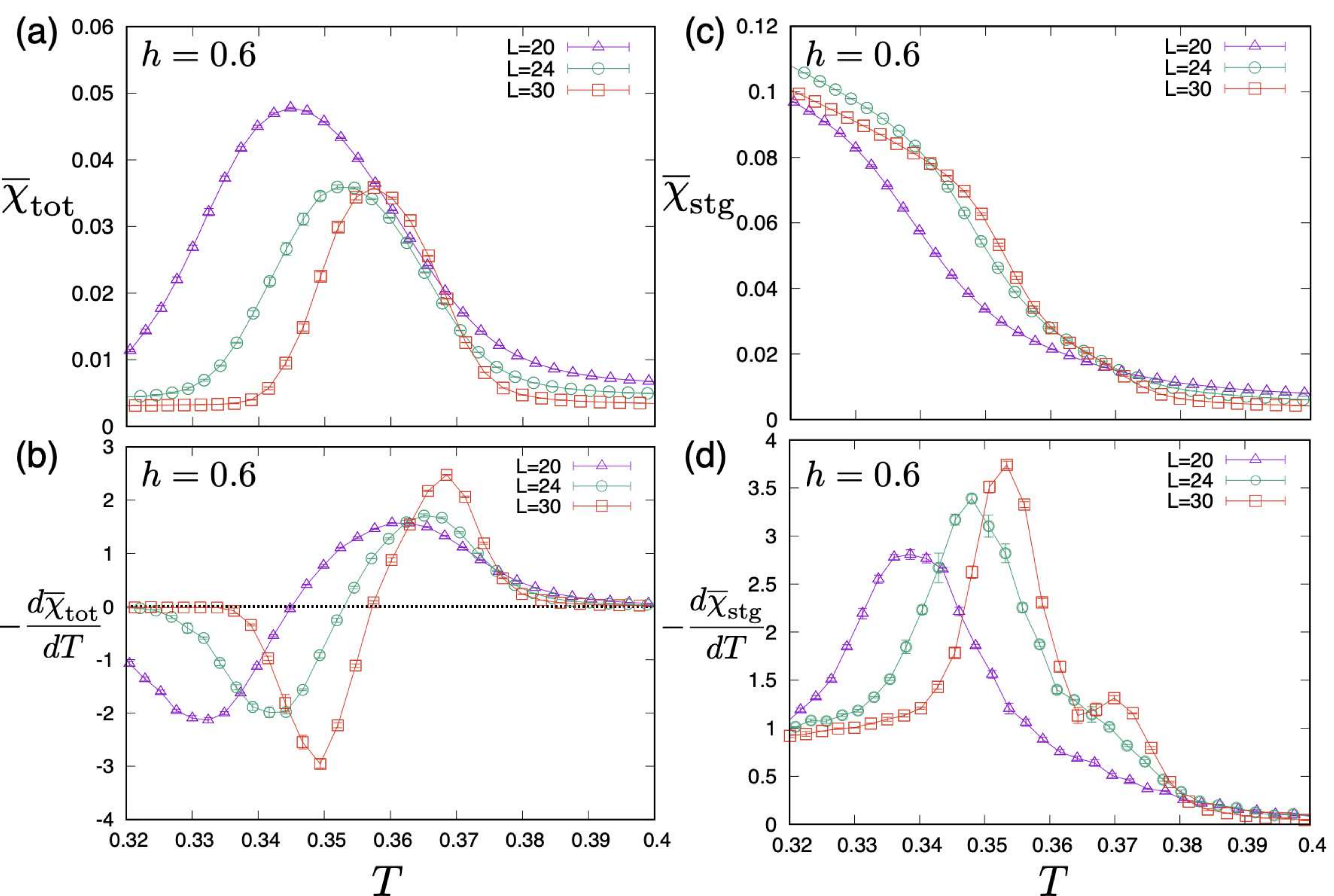}
\caption{The temperature dependence of the chirality-related quantities at a field $h=0.6$: (a) The mean total scalar chirality $\overline{\chi_{\rm tot}}$ and (b) its $T$-derivative, (c) the mean staggered scalar chirality $\overline{\chi_{\rm stg}}$ and (d) its $T$-derivative.
}
\label{chiral_skx}
\end{figure}

 Now, we wish to move on to the more quantitative analysis demonstrating that each constituent state of the RSB SkX phase, i.e., the single-$q$ state and the triple-$q$ SkX state, are magnetically ordered states with the 3D long-range order, and both states macroscopically coexist leading to the RSB. For this purpose, we introduce the order parameters characterizing each constituent state, the total and the staggered scalar chiralities, $\chi_{{\rm tot}}$ and $\chi_{{\rm stg}}$. In terms of the local spin scalar chirality $ \chi_\bigtriangleup$ ($\chi_\bigtriangledown$) defined for the three spins $i,j,k$ located at the corner of the upward (downward) triangle, $\chi_{{\rm tot}}$ and $\chi_{{\rm stg}}$ are defined by
\begin{eqnarray}
\chi_{\rm tot} &=& \frac{1}{2N} \left( \sum_{\bigtriangleup} \chi_{\bigtriangleup} + \sum_{\bigtriangledown} \chi_{\bigtriangledown} \right), \\
\chi_{\rm stg} &=& \frac{1}{2N} \left(\sum_{\bigtriangleup} \chi_{\bigtriangleup} - \sum_{\bigtriangledown} \chi_{\bigtriangledown} \right),
\end{eqnarray}
where the summation $\sum_\bigtriangleup$ ($\sum_\bigtriangledown$) runs over all upward (downward) triangles on the stacked-triangular lattice. These two quantities play complementary roles, i.e. the SkX state exhibits a nonzero total scalar chirality with a vanishing staggered scalar chirality, while the single-$q$ state exhibits a nonzero staggered scalar chirality with a vanishing total scalar chirality. Since the scalar chirality changes its sign under the $Z_2$ spin-mirroring operation, both $\chi_{{\rm tot}}$ and $\chi_{{\rm stag}}$ can be indicators of the $Z_2$ symmetry breaking.

 In Figs. \ref{dist_skx} (a) and (b), we show the distribution of the total chirality, $P(\chi_{{\rm tot}})$, and that of the staggered chirality, $P(\chi_{{\rm stg}})$, respectively, for the sizes $L=20$, 24 and 30 at ($T=0.36$, $h=0.6$) located in the midst of the RSB triple-$q$ SkX phase. As can be seen from Fig. \ref{dist_skx} (a), $P(\chi_{{\rm tot}})$ exhibits three peaks, one zero-peak at $\chi_{{\rm tot}}=0$ corresponding to the single-$q$ state, and the other two side-peaks at symmetric nonzero $\chi_{{\rm tot}}=\pm \chi_{{\rm tot}}^*$ corresponding to the SkX and the anti-SkX states. Interestingly, as the system size $L$ is increased, all three peaks tend to grow and sharpen, suggesting that all three states, i.e., the single-$q$ state, the SkX state and the anti-SkX state, form the genuine ordered state in the thermodynamic limit. By contrast, the weight of the distribution in the intermediate range between 0 and $\pm \chi_{{\rm tot}}^*$ tends to be suppressed with increasing $L$ almost vanishing for $L=30$, indicating that the domain state consisting of finite-size domains is not formed for larger sizes. Hence, the data of $P(\chi_{{\rm tot}})$ indicates the occurrence of the RSB.

 Likewise, as can be seen from Fig. \ref{dist_skx} (b), $P(\chi_{{\rm stg}})$ also exhibits three peaks, one at $\chi_{{\rm stg}}=0$ corresponding to the SkX or the anti-SkX state, and the other two at symmetric nonzero $\chi_{{\rm stg}}=\pm \chi_{{\rm stg}}^*$ corresponding to the single-$q$ states with mutually opposite scalar chiralities. As in the $P(\chi_{{\rm tot}})$ case, all three peaks tend to grow and sharpen as the system size $L$ is increased, whereas the weight of the distribution between the peaks tends to be suppressed. Hence, the data of $P(\chi_{{\rm stg}})$ are fully consistent with those of $P(\chi_{{\rm tot}})$, indicating the occurrence of the RSB. 

 One may suspect that the observed three-peaks structure of the chirality distribution might be due to the possible first-order transition where the coexistence of the two different ordered states is to be expected in equilibrium just at the transition point. This seems rather unlikely here, however, since the three-peak structure of the distribution function is observed in full equilibrium over a rather wide temperature range corresponding to the RSB SkX phase. To further examine the possible relevance of the first-order transition, we show in Fig. \ref{dist_skx} (c) the distribution of the energy per spin, $P(E)$, at ($T=0.3$, $h=0.6$), the same point as that of Figs. \ref{dist_skx} (a) and (b). In contrast to $P(E)$ at the first-order transition point exhibiting a double-peak structure, only a single peak is observed. This observation rules out the possibility that the multiple-peak structure observed in $P(\chi_{{\rm tot}})$ and $P(\chi_{{\rm stg}}$) is originated from the macroscopic phase coexistence associated with a first-order transition.

 The temperature dependence of $P(\chi_{{\rm tot}})$ and $P(\chi_{{\rm stg}})$ is shown in Fig. \ref{dist_skx2} for the size $L=30$, in a wide temperature range covering the paramagnetic phase, the RSB SkX phase  and the single-$q$ phase. In the paramagnetic phase, both $P(\chi_{{\rm tot}})$ and $P(\chi_{{\rm stg}})$ exhibit only a central peak, while in the RSB SkX phase they exhibit the three peaks corresponding to the RSB as mentioned above. As the temperature is lowered within the RSB SkX phase, the central peak grows and the side peaks shrink in $P(\chi_{{\rm tot}})$, while the side peaks grow and the central peak shrinks in $P(\chi_{{\rm stg}})$. At the lowest temperature shown in Fig. \ref{dist_skx2} $T=0.334$, only the central peak (the side peaks) remains in $P(\chi_{{\rm tot}})$ (in $P(\chi_{{\rm stg}})$), indicating that the system is in the single-$q$ state with nonzero $\chi_{{\rm stg}}$ and vanishing $\chi_{{\rm tot}}$.

 We also compute the mean total and staggered scalar chiralities, $\overline{\chi_{\rm tot}} = \sqrt{<\chi_{\rm tot}^2>}$ and $\overline{\chi_{\rm stg}} = \sqrt{<\chi_{\rm stg}^2>}$ together with their $T$-derivatives, $-\frac{{\rm d}\overline{\chi}_{\rm tot}}{{\rm d}T}$ and $-\frac{{\rm d}\overline{\chi}_{\rm stg}}{{\rm d}T}$. Their temperature dependence is shown in Fig. \ref{chiral_skx} for various sizes $L$ for the total chirality (a, b), and for the staggered chirality (c, d). On decreasing the temperature, the mean total chirality increases around $T_{{\rm c1}}$, but decreases around $T_{{\rm c2}}$. One point to be noticed here is that in the $T$-range of $0.358\lesssim T\lesssim 0.369$ $\overline{\chi_{\rm tot}}$ exhibits almost negligible size dependence, suggesting that $\overline{\chi_{\rm tot}}$ remains nonzero in this $T$-range even in the thermodynamic limit. This observation gives an upper bound of $T_{{\rm c2}}$ as $T_{{\rm c2}}\lesssim 0.358$. By contrast, the mean staggered chirality $\chi_{{\rm stg}}$ monotonically increases with decreasing $T$. A closer inspection of the large-size data reveals that this increase occurs in two steps, one around $T_{{\rm c1}}$ and the other around $T_{{\rm c2}}$: See the $L=30$ data of Fig. \ref{chiral_skx} (d). These features are fully consistent with the behavior of the chirality distribution functions shown in Fig. \ref{dist_skx2}.

\begin{figure}[t]
\includegraphics[clip,width=85mm]{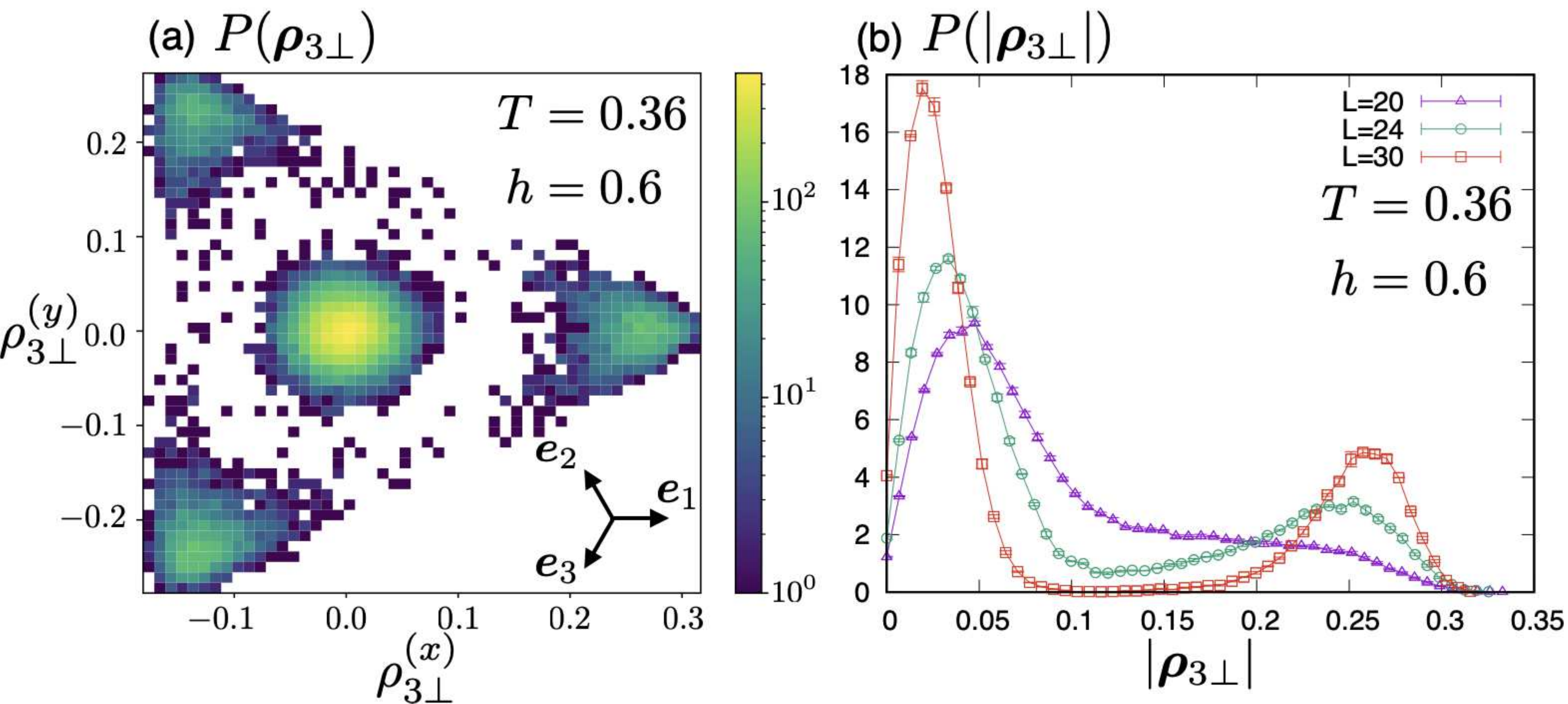}
\caption{The distribution functions of the $Z_3$-symmetry-breaking parameter in the RSB SkX phase at ($h=0.6, T=0.36$) for the size $L=30$: (a) The 2D distribution of the $Z_3$-symmetry breaking parameter $\bm{\rho}_{3\perp}$, and (b) the distribution of the absolute value $|\bm{\rho}_{3\perp}|$ for various sizes $L$.
}
\label{dist_skx3}
\end{figure}

\begin{figure}[t]
\includegraphics[clip,width=85mm]{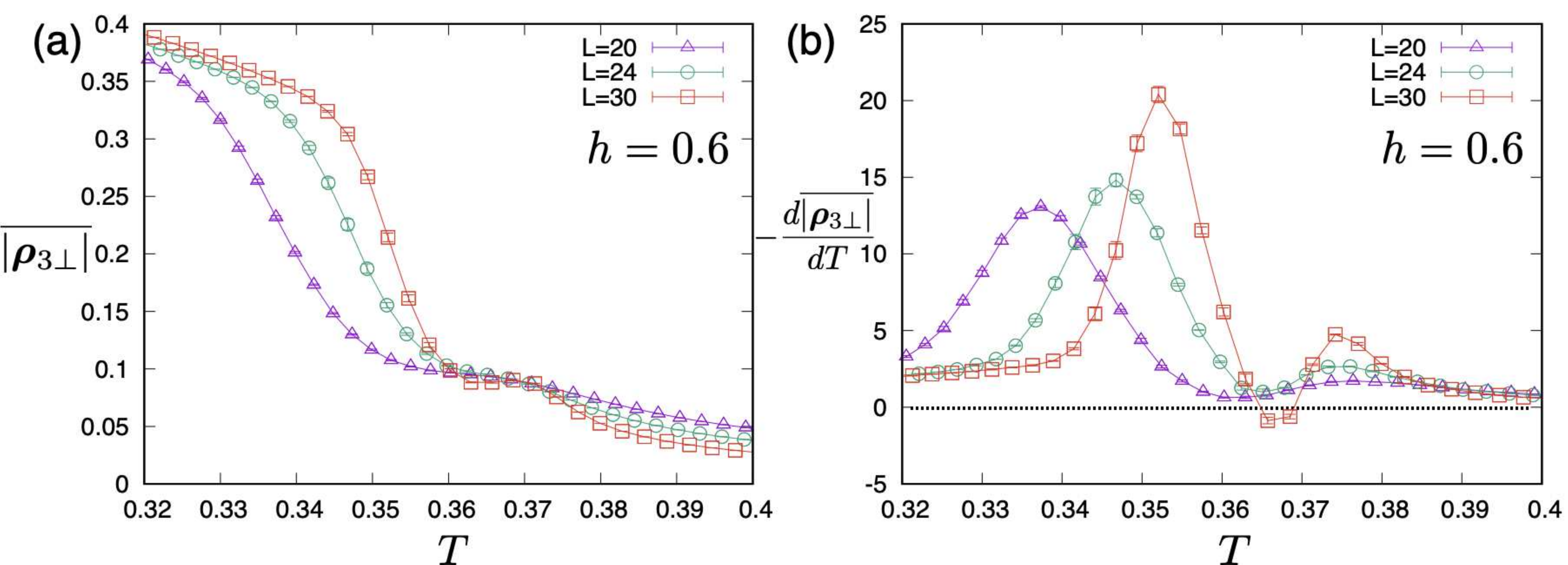}
\caption{The temperature dependence of (a) the mean absolute value of the $Z_3$-symmetry-breaking parameter $\overline{|\bm{\rho}_{3\perp}|}$, and (b) its $T$-derivative. The magnetic field is $h=0.6$
}
\label{rho3v_skx}
\end{figure}

 To further examine the nature of the RSB SkX phase, we introduce the three-fold rotational symmetry breaking order parameter, the $Z_3$-symmetry-breaking parameter, associated with both the perpendicular and the parallel spin components with respect to the magnetic field,
\begin{eqnarray}
\bm{\rho}_{3\perp} &= s_\perp (\bm{q}_1^*) \bm{e}_1 +  s_\perp (\bm{q}_2^*) \bm{e}_2 + s_\perp (\bm{q}_3^*) \bm{e}_3, \label{rho3v} \\
\bm{\rho}_{3\parallel} &= s_\parallel (\bm{q}_1^*) \bm{e}_1 +  s_\parallel (\bm{q}_2^*) \bm{e}_2 + s_\parallel (\bm{q}_3^*) \bm{e}_3 \label{rho3p} ,
\end{eqnarray}
where $\bm{e}_1 = (1,0)$ and $\bm{e}_2 = (-1/2,\sqrt{3}/2)$ are the unit vectors, $s_\perp (\bm{q})$ and $s_\parallel (\bm{q})$ are the instantaneous spin structure factors given by Eqs. (\ref{insta_perp}) and (\ref{insta_para}) without the thermal average. These order parameters $\bm{\rho}_{3\perp}$ and $\bm{\rho}_{3\parallel}$ are two-component vectors defined in the two-dimensional triangular order-parameter space, and represent the extent of the $Z_3$-symmetry breaking. Note that the $Z_3$ symmetry is kept in the SkX state while it is broken in the single-$q$ state. More precisely, this is reflected in the perpendicular component, i.e., $|\bm{\rho}_{3\perp}| = 0$ for the SkX state and $|\bm{\rho}_{3\perp}| > 0$ for the single-$q$ state, whereas, for the parallel spin component, the $Z_3$ symmetry is held both in the SkX and the single-$q$ states, i.e., $|\bm{\rho}_{3\perp}| = 0$ for the both states.

Fig. \ref{dist_skx3} (a) exhibits the two-dimensional distribution of $\bm{\rho}_{3\perp}$ in the midst of the RSB SkX phase at ($T = 0.36$, $h=0.6$) obtained by the fully thermalized replica-exchange MC simulation for $L=30$. The intensity at the origin corresponds to the SkX state preserving the $Z_3$ symmetry, while the intensities at the corner of the triangle correspond to the single-$q$ states, each spot corresponding to the three different broken patterns of the lattice $Z_3$ symmetry. The coexistence of these intensities also supports the RSB picture. 

 Fig. \ref{dist_skx3} (b) exhibits the distribution function of the absolute value of $\bm{\rho}_{3\perp}$, $|\bm{\rho}_{3\perp}|$, for the sizes $L=20$, 24 and 30. The double peak is observed, the peak at smaller $|\bm{\rho}_{3\perp}|$ corresponding to the SkX state and the one  at larger $|\bm{\rho}_{3\perp}|$ to the single-$q$ state. Both peaks tend to grow and sharpen with increasing $L$, suggesting that both states develop into the ordered states in the thermodynamic limit, again supporting the RSB picture.
 
 We also compute the mean $\overline{|\bm{\rho}_{3\perp}|}$ and its $T$-derivative, and their temperature dependence is shown in Figs. \ref{rho3v_skx} (a) and (b) for the sizes $L=20$, 24 and 30. Here, the two-step increase is visible rather clearly, which is more enhanced with increasing the lattice size $L$ as can be seen from Fig. \ref{rho3v_skx} (b). In the temperature range of $0.360\lesssim T\lesssim 0.372$, $\overline{|\bm{\rho}_{3\perp}|}$ exhibits almost negligible size dependence at nonzero values, suggesting that the ordered state is not just the $Z_3$-symmetric pure SkX state, further corroborating the RSB picture.

 The RSB we have identified in the RSB SkX phase has a unique feature that the Hamiltonian and the relevant states are all regular ones: The Hamiltonian is uniform, and the constituent states are all spatially periodic. This is in sharp contrast to the standard RSB cases so far studied, where either the Hamiltonian is random as in spin glasses, or the constituent states are random or glassy even if the Hamiltonian is regular as in molecular or structural glasses.

 Now, we wish to make use of such a unique feature of the present RSB, to illustrate the nature of the RSB in an intuitive and concrete manner. In this procedure, we pay attention to the {\it discrete\/} symmetries of the present mode, i.e., the $Z_2$ symmetry associated with the spin mirroring,  and the $Z_3$ symmetry associated with the lattice rotation. Both $Z_2$ and $Z_3$ symmetries are spontaneously broken in the single-$q$ spiral state, while only the $Z_2$ symmetry is spontaneously broken in the triple-$q$ SkX state. In addition to these discrete symmetries, the model also possesses continuous symmetries, e.g., the U(1) symmetry associated with the spin rotation around the $S_z$-axis and the lattice translation symmetry. These continuous symmetries are also broken spontaneously both in the single-$q$ and the SkX states.

\begin{figure}[t]
\includegraphics[clip,width=85mm]{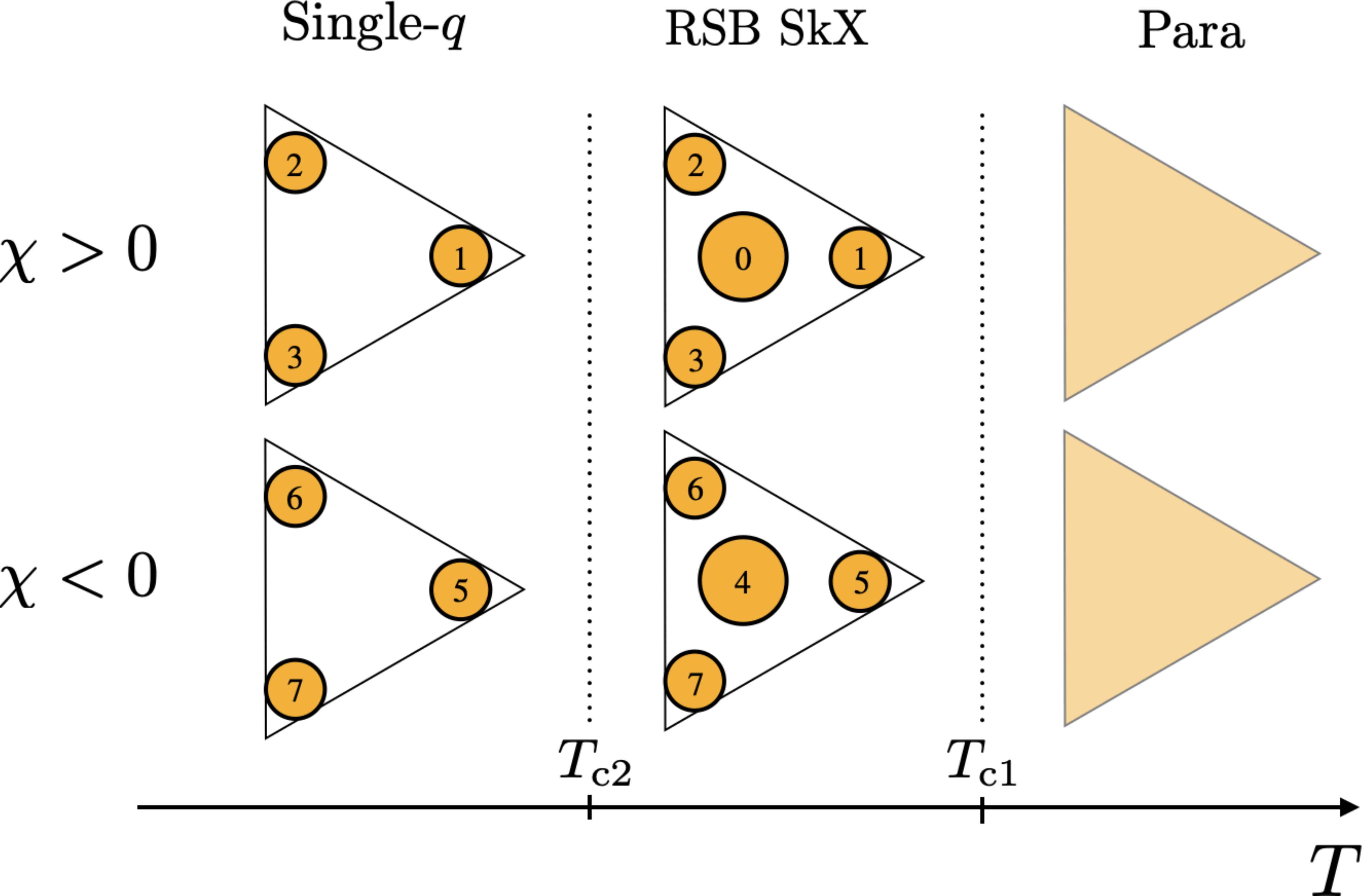}
\caption{Schematic illustration of the phase-space structure after Fig. \ref{dist_skx3} (a) in each paramagnetic phase, the RSB SkX phase and the single-$q$ phase realized with varying the temperature in an applied field. The upper row represents the states with the positive chirality, while the lower row represents the states with the negative chirality. The state 0 and 4 correspond to the triple-$q$ SkX states, while the states (1, 2, 3) and (5, 6, 7) correspond to the single-$q$ states.
}
\label{schematic_rsb}
\end{figure}

 Fig. \ref{schematic_rsb} shows the schematic picture of the phase space after Fig. \ref{dist_skx3} (a) and the symmetry-breaking patterns of the present model. In the paramagnetic phase at $T>T_{{\rm c1}}$, the ergodicity is not broken and the system can visit the entire phase space. On decreasing the temperature beyond $T_{{\rm c1}}$, the system enters the RSB SkX phase where the triple-$q$ SkX state and the single-$q$ spiral state macroscopically coexist. The phase space is divided into eight valleys (pure states): Six of them (1, 2, 3, 5, 6, 7) are the single-$q$ states and two of them (0, 4) are the SkX states. The single-$q$ states (1, 2, 3) can be transformed with each other via the $Z_3$ symmetry operation, similarly for (4, 5, 6), whereas the states (1, 5) can be transformed with each other via the $Z_2$ symmetry operation, and similarly for (2, 6) and (3, 7). The triple-$q$ SkX states (0, 4) keeping the $Z_3$-symmetry can be transformed with each other via the $Z_2$-symmetry operation. In sharp contrast, the single-$q$ states (1, 2, 3, 5, 6, 7) and the triple-$q$ SkX states (0, 4) cannot be transformed by any global symmetry operation of the Hamiltonian. Still, the two sets of states, i.e., the single-$q$ states (1, 2, 3, 5, 6, 7) and the triple-$q$ SkX states (0, 4) macroscopically coexist in the RSB SkX phase at $T_{{\rm c2}}<T<T_{{\rm c1}}$. On further decreasing the temperature beyond $T_{{\rm c2}}$, the weight of the SkX state becomes zero at $T<T_{{\rm c2}}$, and the system becomes replica symmetric in the single-$q$ phase.

\begin{figure}[t]
\includegraphics[clip,width=85mm]{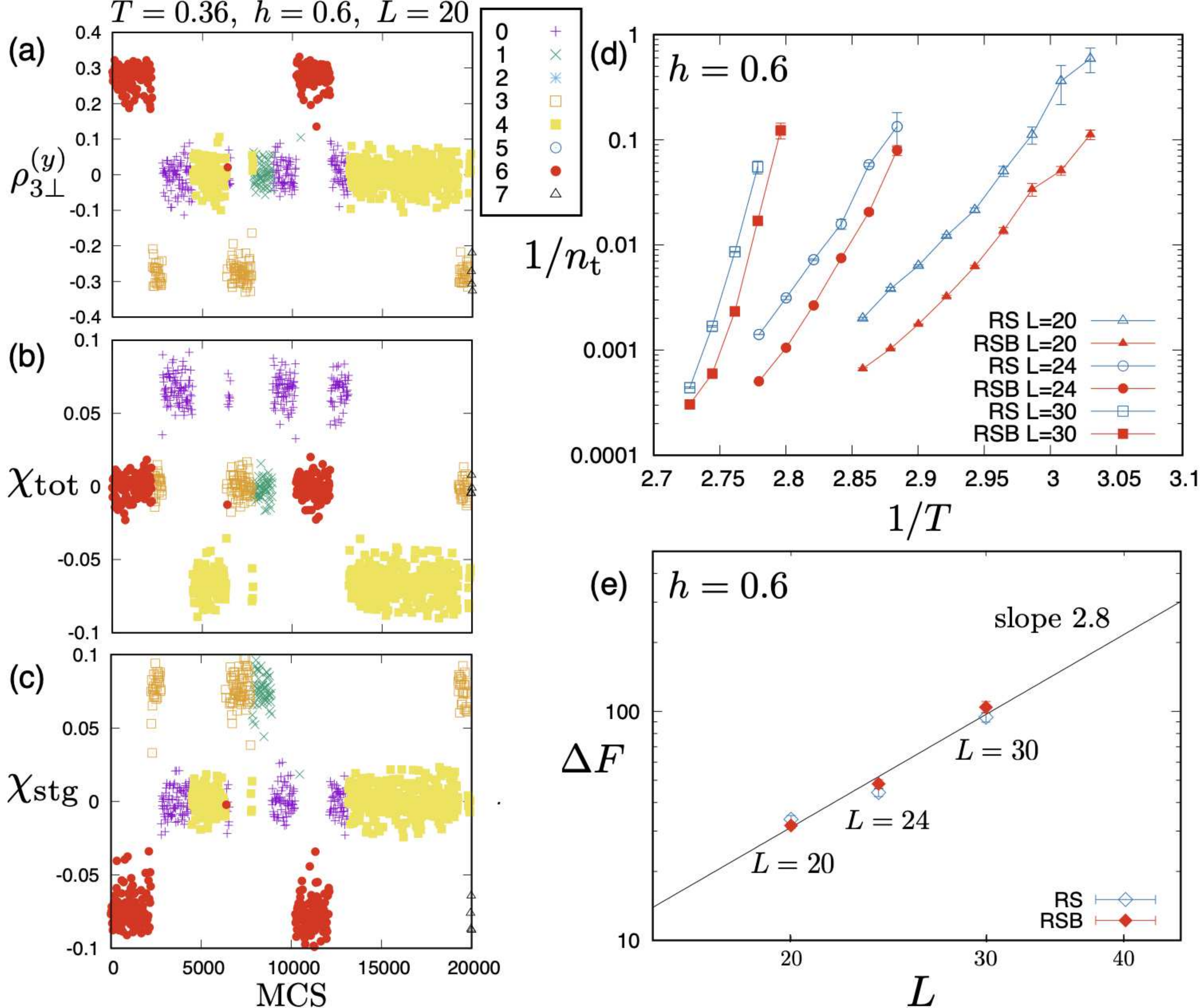}
\caption{The MC time evolutions of (a) the $y$-component of the $Z_3$-symmetry-breaking parameter $\rho_{3\perp}^{(y)}$, (b) the total scalar chirality $\chi_{{\rm tot}}$, and (c) the staggered scalar chirality $\chi_{{\rm stg}}$, in the RSB SkX phase at ($T=0.36, h=0.6$) for the size $L=20$. The color of the data point represents the ordered-state type given in Fig. \ref{schematic_rsb} (a). (d) Arrhenius plot of the state-transition probability for various sizes plotted versus the inverse temperature for each of the RS and the RSB state transitions in the RSB SkX phase at $h=0.6$. (e) The size dependence of the free energy barrier $\Delta F$ estimated from the slope of the data in Fig. (d) on a log-log plot. The straight-line fit gives a slope $\simeq 2.8$ both for the RS and the RSB state transitions.
}
\label{dynamics}
\end{figure}

\subsection{The dynamical simulation of the RSB SkX phase}

 In this section, we continue our study on the RSB SkX phase, but by employing the MC method slightly different from the one employed in the previous subsection, by means of the dynamical simulation. After reaching thermal equilibrium, we turn off the replica-exchange process, and perform MC simulations consisting of the Metropolis and the over-relaxation updates only. Interestingly, even without the replica-exchange process, the jump between different pure states, which we call ``state transition'' below, occurs as a finite-size effect including the one between the single-$q$ and the triple-$q$ SkX states. We then monitor the frequency of each type of state transition to estimate the free-energy barrier between these states, and examine its size dependence.

 Figs. \ref{dynamics} (a)-(c) exhibit the time evolution of the $y$-component of $\bm{\rho}_{3\perp}$, $\rho_{3\perp}^{(y)}$, the total scalar chirality, $\chi_{\rm tot}$, and the staggered scalar chirality, $\chi_{\rm stg}$, during the course of our dynamical simulation performed in the triple-$q$ SkX state at ($T=0.36$, $h=0.6$). As can be seen from the figure, state transitions due to the finite size effect are occasionally observed. These state transitions can be classified into two categories: The one is the transition between the states transformable via by the $Z_2$ or $Z_3$ global symmetry operation, and the other is the transition between the states not transformable via the $Z_2$ or $Z_3$ symmetry operation, i.e. the transition between the single-$q$ and the triple-$q$ SkX states. We call the former ``the RS state transition'' and the latter ``the RSB state transition''. We then measure each state transition frequency $n_t$, which is expected to be inversely proportional to the average time to overcome the free energy barrier separating the pure states. Fig. \ref{dynamics} (d) shows the inverse frequencies $1/n_{\rm t}$ of the RS and RSB state transitions for $L=20,24,30$ plotted versus the inverse temperature $1/T$. We find that the inverse frequencies of both the RS and RSB transitions follow the Arrhenius law,
\begin{equation}
\frac{1}{n_{\rm t}} \sim \exp \left( -\frac{\Delta F}{T} \right) ,
\label{arrhe}
\end{equation}
which enables us to estimate $\Delta F$ from the slope of the data given in Fig. \ref{dynamics} (d). In Fig. \ref{dynamics} (e), the free energy barrier $\Delta F$ obtained in this way are plotted versus the system size $L$  on a log-log plot. $\Delta F$ of both the RS and the RSB state transitions turn out to be well fitted by the power-law $\approx L^\alpha$ with a common exponent $\alpha \simeq 2.8$, indicating that the free energy barrier diverges in the thermodynamic limit and the ergodicity is broken, not only in the RS case but also in the RSB case. This gives another independent support of the RSB occurring in this phase.

 

\begin{figure}[t]
\includegraphics[clip,width=87mm]{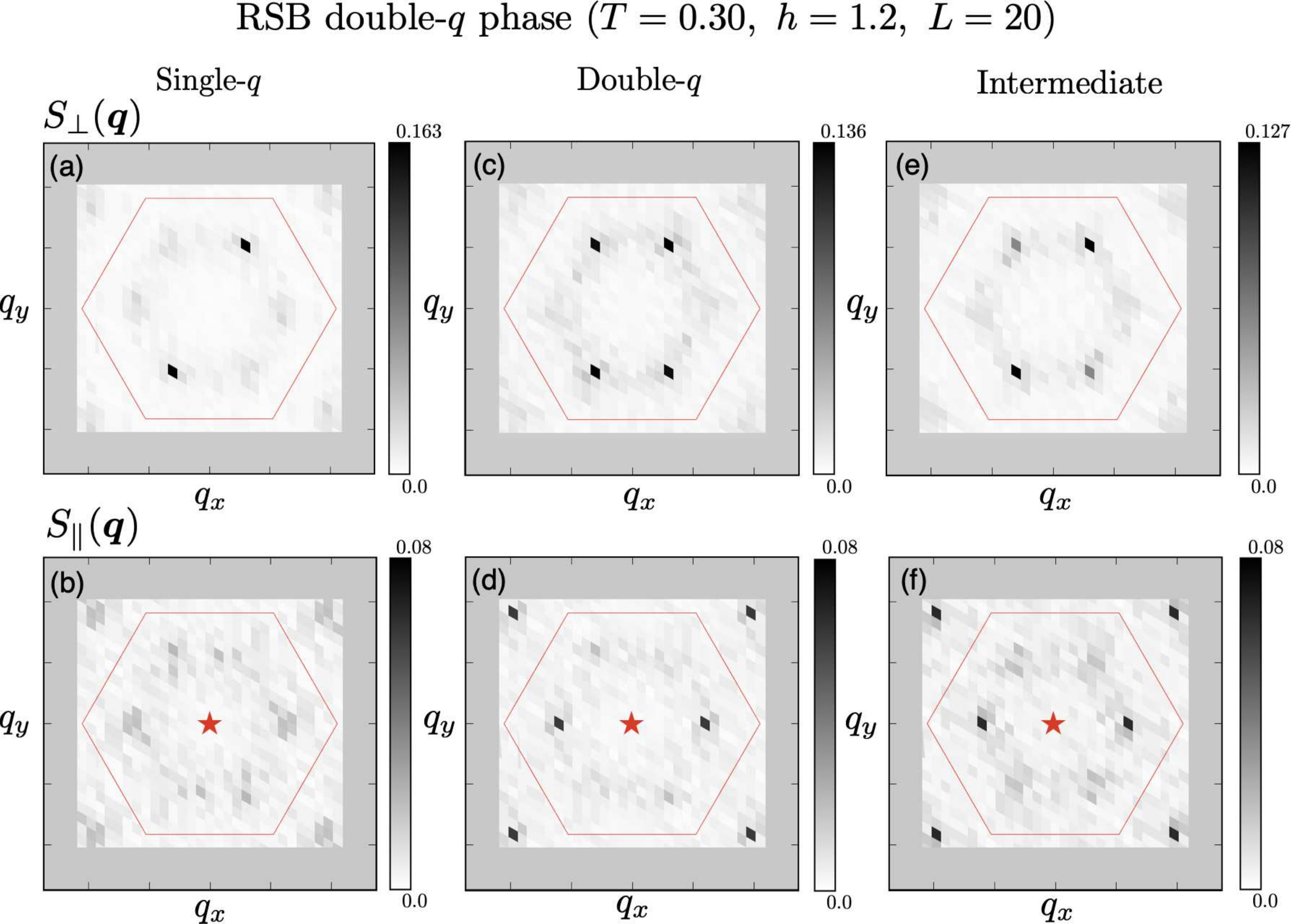}
\caption{The perpendicular spin structure factor $S_{\perp}(\bm{q})$ [the upper row; (a), (c) and (e)] and the parallel spin structure factor $S_{\parallel}(\bm{q})$ [the lower row; (b), (d) and (f)] in the $q_z = 0$ plane realized as an equilibrium state in the RSB double-$q$ phase at ($T=0.30, h=1.2$) for the size $L=20$. After reaching thermal equilibrium, the replica-exchange and the over-relaxation processes are turned off, and the measurements are made by making the short-time average of 100 MCS. The red hexagon represents the first Brillouin zone, and the red star at the origin represents the intensive uniform $\bm{q} = \bm{0}$-component induced by applied magnetic fields: The left column (a) and (b) corresponds to the single-$q$ state, the middle column (c) and (d) to the double-$q$ state, and the right column (e) and (f) to the intermediate state.
}
\label{snap_double3}
\end{figure}

\begin{figure}[t]
\includegraphics[clip,width=85mm]{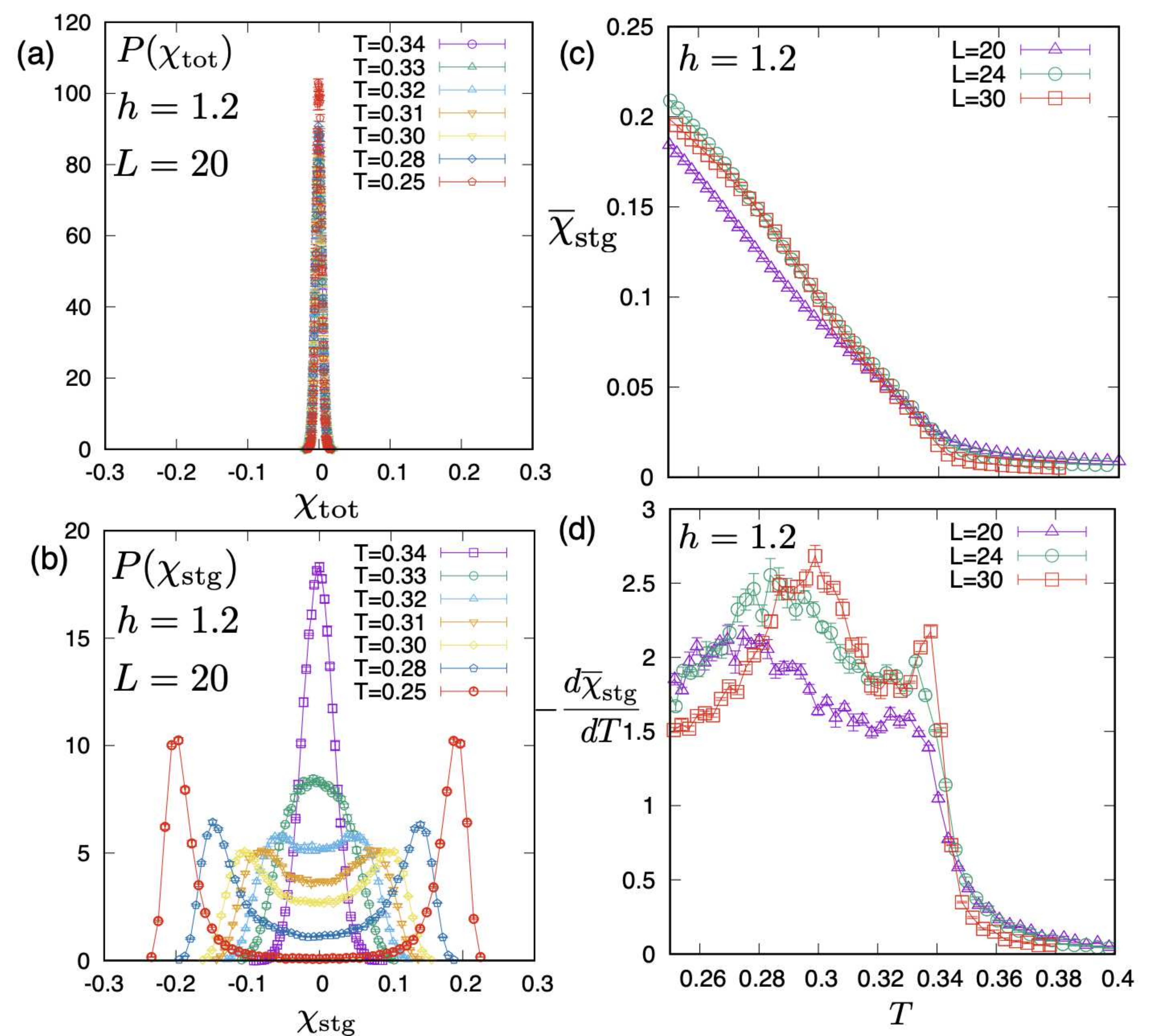}
\caption{The temperature dependence of equilibrium chirality distribution functions at a magnetic filed $h=1.2$ at various temperatures, of (a) the total scalar chirality, and of (b) the staggered scalar chirality. The lattice size is $L=20$. The temperature dependence of (c) the mean staggered chirality $\overline{\chi}_{{\rm stg}}$, and of (d) its $T$-derivative for various sizes.
}
\label{chiral_double}
\end{figure}

\begin{figure}[t]
\includegraphics[clip,width=85mm]{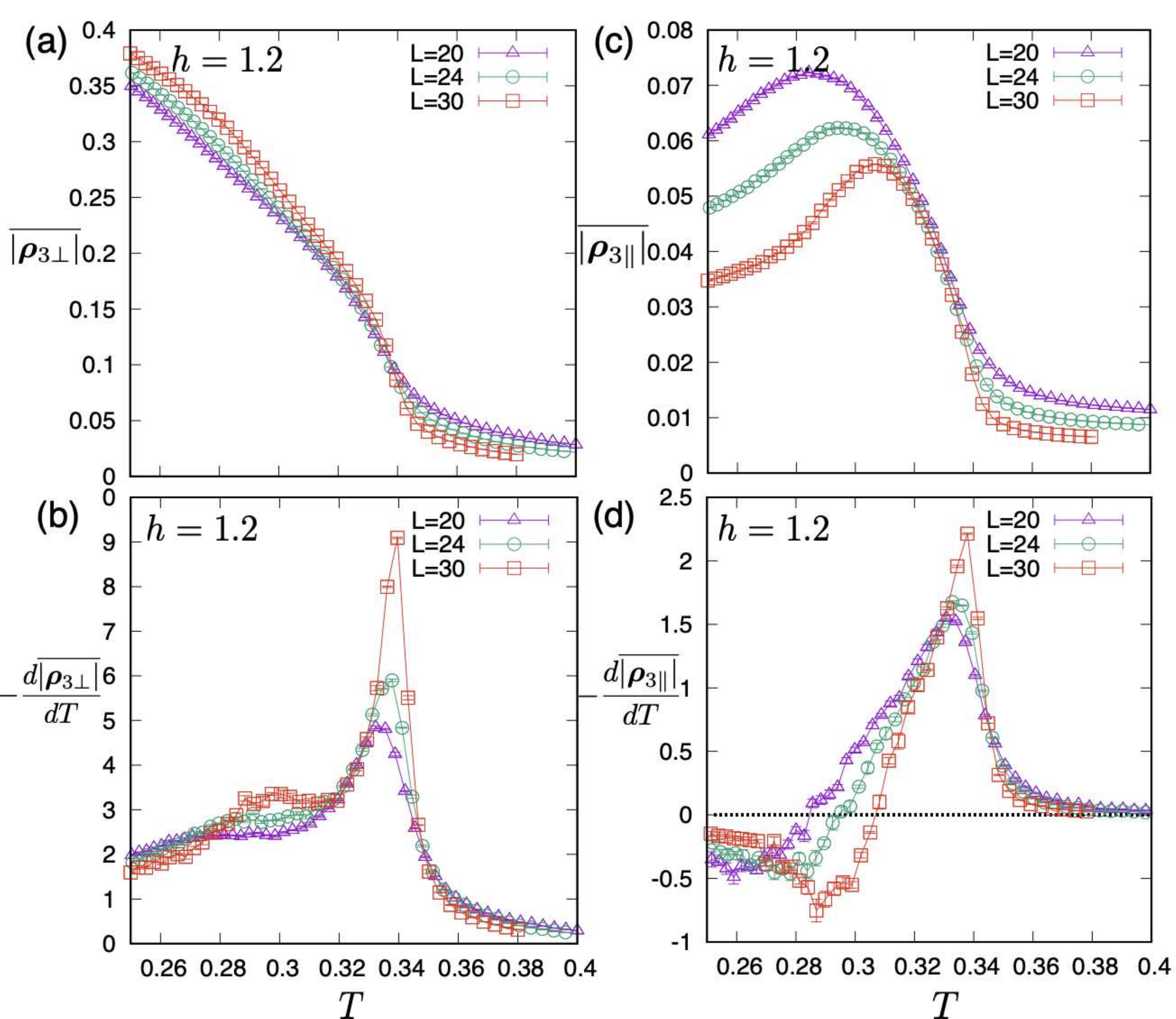}
\caption{The temperature dependence of the $Z_3$-symmetry-breaking parameters at a magnetic field $h=1.2$ for various sizes. (a) The mean absolute value of the perpendicular one $\overline{|\bm{\rho}_{3 \perp}|}$, and (b) its $T$-derivative. (c) The mean absolute value of the parallel one $\overline{|\rho_{3 \parallel}|}$, and (d) its $T$-derivative.
}
\label{physical_double}
\end{figure}

\subsection{The RSB double-$q$ phase}

 In this subsection, we study the nature of the RSB double-$q$ phase stabilized in the high-field region of the phase diagram shown in Fig. \ref{phase}. If one recalls that, in the case of the 2D short-range model studied earlier, the corresponding part of the phase diagram is occupied by the double-$q$ phase \cite{okubo2012multiple}, natural expectation might be that the stable phase in this high-field region is the double-$q$ phase. In the double-$q$ state, both the total and the staggered scalar chiralities, $\chi_{{\rm tot}}$ and $\chi_{{\rm stg}}$, vanish: See below and Appendix \ref{sec_double_spin_chiral}.

 In this subsection, we fix the magnetic field to a typical value $h=1.2$, and employ the fully thermalized MC simulation with the replica-exchange process. The specific-heat data shown in Fig. \ref{heatcap} (b) shows that, at this field of $h=1.2$, a phase transition from the paramagnetic phase to the ordered phase takes place at $T=T_{{\rm c1}}\simeq 0.34$.

 To probe the nature of the ordered state below $T_{{\rm c1}}$, we compute the spin structure factor both for the perpendicular and parallel components, $S_\perp({\bm q})$ and $S_\parallel({\bm q})$, at $T=0.30$, and the results are shown in Fig. \ref{snap_double3}. As in the case of Figs. \ref{snap_skx} (c) and (d) of subsection A, we first fully thermalize the system by utilizing the combination of the Metropolis, the over-relaxation and the replica-exchange processes. Such fully thermalized states turn out to provide a variety of patterns of $S_\perp({\bm q})$ and $S_\parallel({\bm q})$ as the MC time goes on. We then turn off the over-relaxation and the replica exchange process at a certain point to ``pick up'' each pattern, and compute both $S_\perp({\bm q})$ and $S_\parallel({\bm q})$ by the short-time averaging of 100 MCS consisting of the Metropolis updating only to reduce the thermal noise. Three typical $S({\bm q})$ patterns obtained in this way are given in Fig. \ref{snap_double3}, $S_\perp({\bm q})$ in the upper row and $S_\parallel({\bm q})$ in the lower row. One typical pattern is that of the single-$q$ state as shown in Figs. \ref{snap_double3} (a) and (b), and the other typical one is that of the double-$q$ state as shown in Figs. \ref{snap_double3} (c) and (d).

 The double-$q$ state is characterized by the appearance of two pairs of intensities in $S_\perp({\bm q})$ as shown in Fig. \ref{snap_double3} (c), and of one pair of intensities in $S_\parallel({\bm q})$ which is complementary to those in $S_\perp({\bm q})$ as shown in Fig. \ref{snap_double3} (d). The typical spin and chirality configurations of the double-$q$ state in real space are shown in Figs. \ref{double_app2} (a) and (b) of Appendix \ref{sec_double_spin_chiral}. As can be seen from these figures, the $S_z$-component of spin and the scalar chirality form a single-$q$ linear density wave along a certain direction on the triangular lattice, say, the $\bm{q}_1^*$-direction which is complementary to the $\bm{q}_2^*$ and $\bm{q}_3^*$ directions forming the double-$q$ structure in $S_\perp(\bm{q})$: See Figs. \ref{snap_double3} (c) and (d). From Fig. \ref{double_app2} (b) of Appendix \ref{sec_double_spin_chiral}, one can also see that both the total and the staggered chiralities vanish in the double-$q$ state. These features of the double-$q$ state are similar to the double-$q$ state of the short-range $J_1$-$J_3$ ($J_1$-$J_3$-$J_{1c}$) model \cite{okubo2012multiple, osamura}.

 Since both the single-$q$ and the double-$q$ states are realized in equilibrium in this phase, the phase is likely to be the RSB state where the single-$q$ state and the double-$q$ state coexist macroscopically. Thus, we call the state the ``RSB double-$q$ phase''. While the RSB double-$q$ phase is similar to the RSB triple-$q$ SkX phase consisting of the single-$q$ state and the SkX state studied in previous subsections, it also has a different character. As shown in  Figs. \ref{snap_double3} (e) and (f), the intermediate-type state, where the two pairs of $S_\perp({\bm q})$ intensities in the double-$q$ pattern have mutually different intensities. Indeed, such intermediate states with continuously varying relative intensities are observed from one equilibrium state to the other, apparently connecting the pure single-$q$ and the pure double-$q$ states. Hence, the RSB pattern appears to be continuously degenerate unlike the case of the RSB SkX phase.

 In this high-field region, the triple-$q$ SkX state is completely absent. As shown in Fig. \ref{chiral_double} (a), the total chirality distribution is always sharply zero-peaked for $h=1.2$, indicating the absence of the SkX state. In Fig. \ref{chiral_double} (b), we show the staggered chirality distribution $P(\chi_{{\rm stg}})$ for various temperatures at $h=1.2$. Below $T_{{\rm c1}}\simeq 0.34$ determined from the specific-heat peak, $P(\chi_{{\rm stg}})$ gradually develops a nontrivial form with symmetric side peaks located at $\pm \chi_{{\rm stg}}^*$, while the weight of the distribution between the two side peaks remains finite in the wide temperature region below $T_{c1}$, in contrast to the case of the RSB SkX phase shown in Fig. \ref{dist_skx} (b). These nonzero weights are likely to arise from the double-$q$ state and the intermediate state identified in Figs. \ref{snap_double3} (e) and (f). When the temperature is further lowered, the weight between the side peaks eventually goes away. The result suggests that the low-temperature phase below $T_{{\rm c2}}\simeq 0.30$ is the replica-symmetric single-$q$ state.

 We also compute the mean staggered scalar chirality $\overline{\chi_{\rm stg}}$ and its $T$-derivative $-\frac{{\rm d}\overline{\chi_{\rm stg}}}{{\rm d}T}$, and their temperature dependence is shown in Figs. \ref{chiral_double} (c) and (d) for various sizes $L$. Note that the staggered chirality is an indicator of the single-$q$ state, since it becomes zero for the double-$q$ and the triple-$q$ SkX states. As can be seen from the figures, especially from Fig. \ref{chiral_double} (d), the staggered chirality grows in two steps on decreasing the temperature, and this tendency is more eminent for larger sizes. These two-step growth is associated with $T_{{\rm c1}}$ and $T_{{\rm c2}}$, and is consistent with our interpretation that the phase at $T_{{\rm c2}}<T<T_{{\rm c1}}$ is the RSB double-$q$ phase, while the one at $T<T_{{\rm c2}}$ is the standard single-$q$ phase.

 In order to get further insight into the intermediate state apparently connecting the single-$q$ and double-$q$ states, we perform the mean-field analysis based on the Landau theory \cite{reimers1991mean, okubo2012multiple}. The details of the analysis is given in Appendix \ref{sec_mean_app}. The mean-field equation possesses the solution which is a superposition of the single-$q$ and the double-$q$ solutions,
\begin{equation}
\bm{S}_i = \bm{S}_i^{(s)} + \bm{S}_i^{(d)},
\label{single_plus_double}
\end{equation}
where
\begin{equation}
\bm{S}_i^{(s)} = \left(
\begin{array}{c}
I_{xy}^{(s)}\cos(\bm{q}_3^*\cdot \bm{r}_i + \theta_3)  \\
I_{xy}^{(s)}\sin(\bm{q}_3^*\cdot \bm{r}_i + \theta_3)  \\
0
\end{array}
\right)
\label{single}
\end{equation}
\begin{equation}
\bm{S}_i^{(d)} = \left(
\begin{array}{c}
I_{xy}^{(d)} (\cos(\bm{q}_2^*\cdot \bm{r}_i + \theta_2) + \cos(\bm{q}_3^*\cdot \bm{r}_i + \theta_3)) \\
- I_{xy}^{(d)} (\sin(\bm{q}_2^*\cdot \bm{r}_i + \theta_2) + \sin(\bm{q}_3^*\cdot \bm{r}_i + \theta_3)) \\
s_{\parallel}(\bm{0}) + 2s_{\parallel}(\bm{q}_1^*)\cos(\bm{q}_1^*\cdot \bm{r} + \theta_1) 
\end{array}
\right),
\label{double}
\end{equation}
%


where $I_{xy}^{(s)} = \sqrt{2}(s_{\perp}(\bm{q}_3^*) - s_{\perp}(\bm{q}_2^*))$,
$I_{xy}^{(d)} = \sqrt{2}s_{\perp}(\bm{q}_2^*)$, and $\theta_i~(i=1,2,3)$ is an arbitrary phase parameter satisfying $\cos (\theta_1 + \theta_2 + \theta_3) = -1$.

 Indeed, we find that once the parameters contained in the above formula, Eqs. (\ref{single}) and (\ref{double}), are properly chosen, the resulting mean-field spin configurations well reproduce the MC spin configurations of the RSB double-$q$ state. The detailed procedure and the comparison with the MC results are given in Appendix \ref{sec_mean_app}.

 The $Z_3$-symmetry breaking parameters $\bm{\rho}_{3 \perp}$ and $\bm{\rho}_{3 \parallel}$ are also informative. Note that, for the perpendicular spin structure factor $S_\perp(\bm{q})$, the $Z_3$ symmetry is broken both in the single-$q$ and the double-$q$ states, but not in the triple-$q$ SkX state. By contrast, for the parallel spin structure factor $S_\parallel(\bm{q})$, the $Z_3$ symmetry is broken only in the double-$q$ state, not in the single-$q$ nor the triple-$q$ SkX state. As a result, $\bm{\rho}_{3\parallel}$ becomes nonzero only for the double-$q$ state. This means that $|\bm{\rho}_{3\parallel}|$ serves as the order parameter of the double-$q$ ordered state.

 In Fig.\ref{physical_double}, we show the temperature dependence of $\overline{|\bm{\rho}_{3 \perp}|}$ (a) and $\overline{|\bm{\rho}_{3 \parallel}|}$ (c), and their temperature derivative, $\frac{{\rm d}\overline{|{\bm \rho}_{3 \perp}|}}{{\rm d}T}$ (b) and $\frac{{\rm d}\overline{|\bm{\rho}_{3 \parallel}|}}{{\rm d}T}$ (d). The two-step growth of $|\bm{\rho}_{3 \perp}|$ associated with $T_{{\rm c1}}$ and $T_{{\rm c2}}$ is discernible for larger sizes (see the double-peak structure of $\frac{{\rm d}\overline{|\bm{\rho}_{3 \perp}|}}{{\rm d}T}$ shown in Fig. \ref{physical_double} (b)). By contrast, Fig.\ref{physical_double} (c) indicates that $\overline{|\bm{\rho}_{3 \parallel}|}$ exhibits the non-monotonic temperature dependence, remaining nonzero only in the intermediate $T$-range between $T_{{\rm c1}}$ and $T_{{\rm c2}}$. This observation suggests that the double-$q$ state comes into play only between $T_{{\rm c1}}$ and $T_{{\rm c2}}$, consistently with the above identification that the phase  between $T_{{\rm c1}}$ and $T_{{\rm c2}}$ is the RSB double-$q$ phase and the one below $T_{{\rm c2}}$ is the standard single-$q$ phase.

\begin{figure}[t]
\includegraphics[clip,width=87mm]{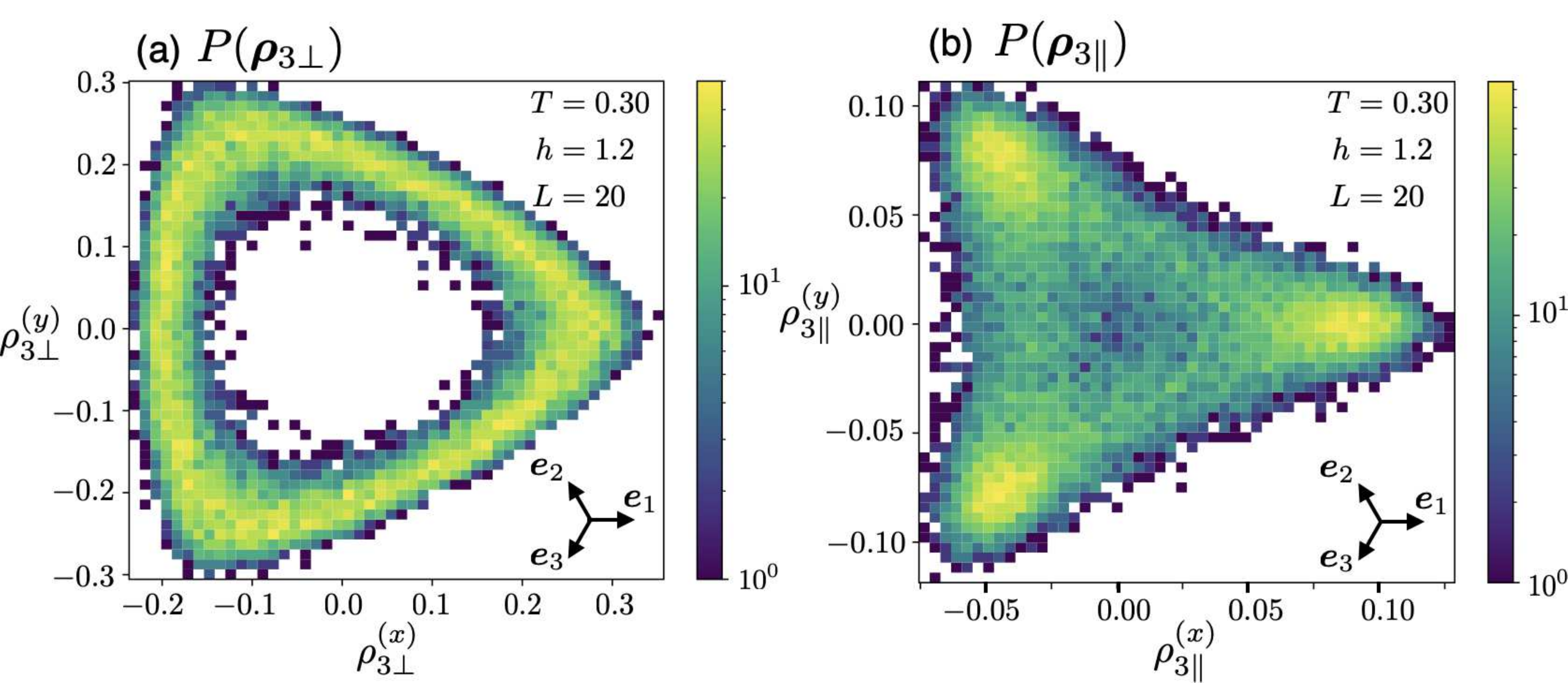}
\caption{
The 2D distribution of the $Z_3$-symmetry-breaking parameters in the RSB double-$q$ phase at ($T=0.30$, $h=1.2$) for (a) the perpendicular component ${\bm \rho}_{3 \perp}$, and for (b) the parallel component ${\bm \rho}_{3 \parallel}$. The lattice size is $L=20$.
}
\label{snap_double}
\end{figure}

 We next investigate the behavior of $\bm{\rho}_\perp$ and  $\bm{\rho}_\parallel$ in the RSB double-$q$ state in the full two-component order parameter space. The typical distributions of $\bm{\rho}_\perp$ and $\bm{ \rho}_\parallel$ are shown in Figs. \ref{snap_double} (a) and (b), respectively, at ($T=0.3$, $h=1.2$). The perpendicular one shown in Fig. \ref{snap_double} (a) exhibits the triangular ring form in the two-dimensional order parameter space without an appreciable weight around the origin. Here, note that the corner of the triangle corresponds to the single-$q$-like $S(\bm{q})$ pattern, the middle of the triangle side to the double-$q$-like $S(\bm{q})$ pattern, and the center (origin) to the triple-$q$-like or the $Z_3$-symmetry-preserved  $S(\bm{q})$ pattern. Fig. \ref{snap_double} (a) demonstrates clearly that the RSB double-$q$ state consists of the single-$q$, the double-$q$ and intermediate states spanning these two states in a continuous manner.

 Likewise, the parallel one $\overline{|\rho_{3\parallel}|}$ shown in Fig. \ref{snap_double} (b) also exhibits the continuous triangular distribution, with enhanced weights at the single-$q$ points and with an appreciable weight around the origin, in contrast to the perpendicular case shown in Fig. \ref{snap_double} (a). This difference reflects the facts that, in the parallel $S_\parallel(\bm{q})$, the double-$q$-like pattern is realized neither in the single-$q$ nor in the double-$q$ state, and that the pure single-$q$ state preserves the $Z_3$ symmetry in $S_\parallel(\bm{q})$.

 Note that the RSB double-$q$ phase observed here is not a floating phase which is known to appear in several low-dimensional models such as the 2D clock model \cite{cardy1980general, tobochnik1982properties, challa1986critical}. This can be confirmed from the observation that the order parameters such as $\chi_{\rm stg},~\bm{\rho}_{3\perp}$ and $\bm{\rho}_{3\parallel}$ are nonzero in the RSB double-$q$ phase in the thermodynamic limit, which means that each constituent state is the long-range ordered state and the ergodicity between the constituent states is broken in the thermodynamic limit.

\begin{figure}[t]
\includegraphics[clip,width=87mm]{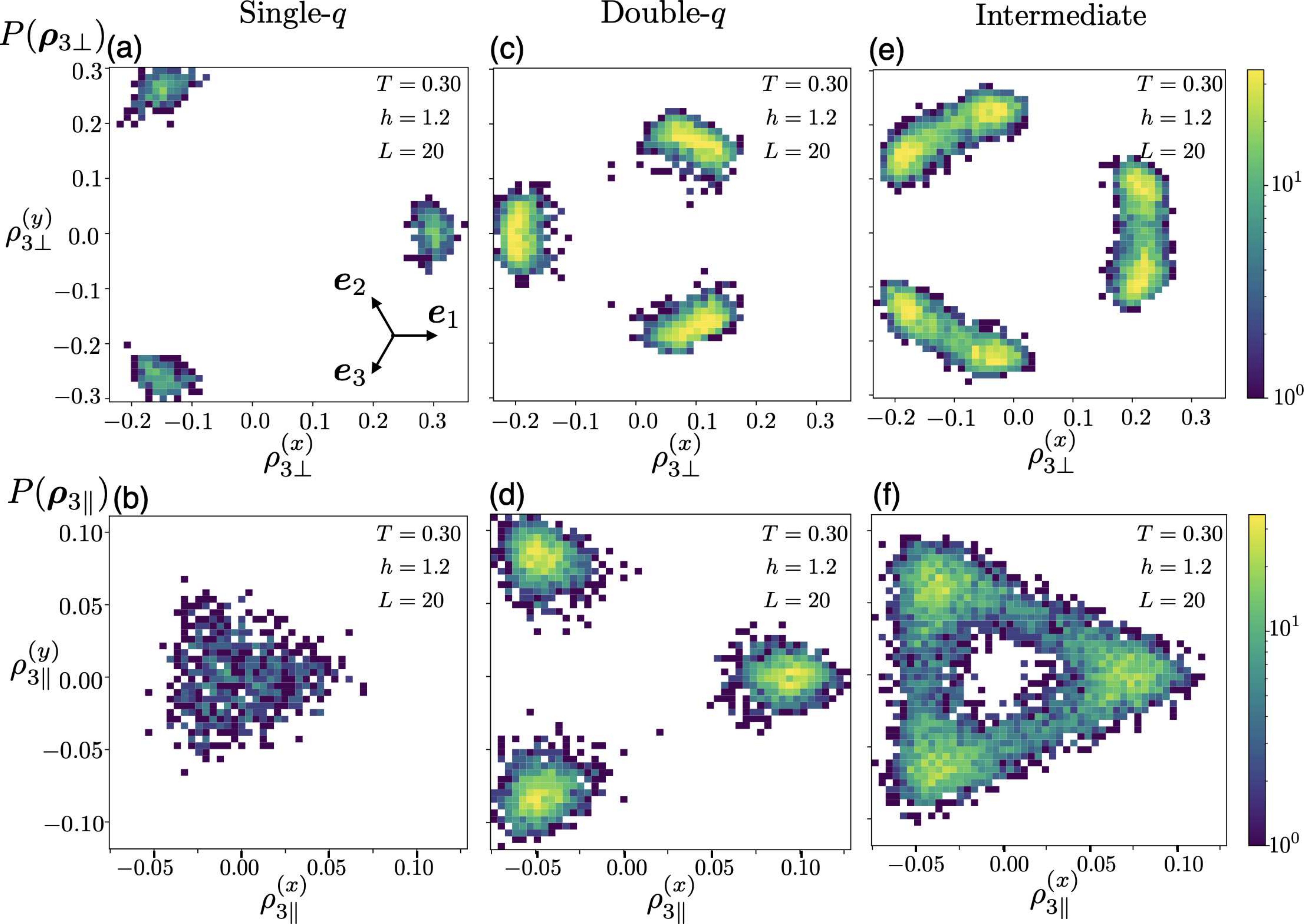}
\caption{
The 2D distribution of the $Z_3$-symmetry-breaking parameters in the RSB double-$q$ phase at ($T=0.30$, $h=1.2$). The lattice size is $L=20$. The upper row corresponds to the perpendicular component ${\bm \rho}_{3 \perp}$ [(a), (c), (e)], and the lower row to the parallel component ${\bm \rho}_{3 \parallel}$ [(b), (d), (f)]. The left column corresponds to the single-$q$-like state satisfying the constraint $4\chi_{\rm stg}^{\rm max}/5 < |\chi_{\rm stg}| < \chi_{\rm stg}^{\rm max}$ [(a), (b)], the middle column to the double-$q$-like state  satisfying the constraint $|\chi_{\rm stg}| < \chi_{\rm stg}^{\rm max}/5$ [(c), (d)], and the right column to the intermediate state satisfying the constraint $2\chi_{\rm stg}^{\rm max}/5 < |\chi_{\rm stg}| < 3\chi_{\rm stg}^{\rm max}/5$ [(e), (f)]. 
}
\label{snap_double2}
\end{figure}

 In order to give further support to our picture of the RSB double-$q$ phase, especially the nature of intermediate states spanning from the single-$q$ state to the double-$q$ state, we investigate the  2D $\bm{\rho}_\perp$ and  $\bm{\rho}_\parallel$ distributions in terms of the staggered-chirality distribution shown in Fig. \ref{chiral_double} (b). In Fig. \ref{snap_double2}, we draw the 2D $\bm{\rho}_\perp$ and  $\bm{\rho}_\parallel$ distributions for the divided subspaces, divided depending on the values of the staggered scalar chirality. More precisely, we divide the staggered-chirality space $0 < |\chi_{\rm stg}| < \chi_{\rm stg}^{\rm max}$ into five equally spaced subregions, and show the sub-averaged ${\bf \rho}_\perp$ and  ${\bf \rho}_\parallel$ distributions (upper row and lower row, respectively), each corresponding to the subregion $4\chi_{\rm stg}^{\rm max}/5 < |\chi_{\rm stg}| < \chi_{\rm stg}^{\rm max}$ [Figs. \ref{snap_double2} (a) and (b)], $|\chi_{\rm stg}| < \chi_{\rm stg}^{\rm max}/5$ [Figs. \ref{snap_double2} (c) and (d)], and $2\chi_{\rm stg}^{\rm max}/5 < |\chi_{\rm stg}| < 3\chi_{\rm stg}^{\rm max}/5$ [Figs. \ref{snap_double2} (e) and (f)]. The results clearly demonstrate the character of each constituent state forming the RSB double-$q$ phase.

\section{summary and discussion}
\label{summary}

 To summarize, we performed extensive equilibrium MC simulations of the RKKY classical Heisenberg model on the 3D stacked-triangular lattice. We determined the magnetic phase diagram of the model in the temperature versus magnetic-field plane for a typical parameter set realizing the incommensurate magnetic order. While the obtained phase diagram turned out to exhibit the phase structure more or less similar to the one of the 2D short-range model, containing the single-$q$, the double-$q$ and the triple-$q$ SkX states, the character of each phase turned out to be entirely different. Namely, we have found that the two of ordered phases, i.e., the RSB double-$q$ phase and the RSB triple-$q$ SkX phase, exhibit the unexpected RSB phenomenon, where the double-$q$ or the triple-$q$ SkX states macroscopically coexist with the single-$q$ state. The free energies of the constituent states are degenerate at $\mathcal{O}(1)$, while the free-energy barrier separating these states diverges in the thermodynamic limit, and the ergodicity is spontaneously broken. Since the single-$q$ state and the triple-$q$ SkX state (or the double-$q$ state) have entirely different symmetries and cannot be transformable by any symmetry operation of the Hamiltonian, the spontaneous symmetry breaking observed here differs from the standard Hamiltonian-symmetry based spontaneous symmetry breaking, i.e., the spontaneous RSB.

 From the viewpoint of discrete symmetries, the triple-$q$ SkX phase consists of a finite number of pure states, only eight states: The two of them are triple-$q$ SkX states related via the $Z_2$ symmetry, and the six of them are single-$q$ states related via the $Z_2$ and $Z_3$ symmetries ($6=2\times 3$), whereas these two and six states cannot be related via any symmetry operation. By contrast, the RSB double-$q$ phase actually consists of an infinite number of states: In addition to the single-$q$ and the double-$q$ states, there is an infinite number of intermediate states which are superpositions of the single-$q$ and the double-$q$ states with arbitrary relative weight, continuously connecting the single-$q$ and the double-$q$ states. Even in such a situation, the ergodicity is broken among an infinite number of constituent states.

 Note that the RSB we identified for the 3D RKKY model is not realized in the 2D short-range model, e.g., the $J_1-J_3$ ($J_1-J_2$) triangular Heisenberg model, where only the standard replica-symmetric triple-$q$ SkX or the double-$q$ state is realized. Which aspect of the model brings about the unique RSB character is an interesting question. There might be two possible driving factors. One is the difference in the spatial dimensionality, 3D in the present case versus 2D in Ref.\cite{okubo2012multiple}, and the other is the difference in the interaction range, the long-range RKKY interaction in the present case versus the short-range interaction in Ref.\cite{okubo2012multiple}. To get further insight, further study on the 3D short-range mode and/or the 2D RKKY model would be informative. 

 We note that the RSB was also not reported in the 2D Kondo-lattice model on the triangular lattice studied in Ref.\cite{ozawa2017zero}. There, the lattice was 2D, and the SkX was stabilized via the competition (or frustration) between the two-spin and the four-spin (biquadratic) interactions \cite{hayami2017effective}, in contrast to the competition between the ferromagnetic and the antiferromagnetic two-spin interactions in the present RKKY model. How the difference in the frustration type affects (or does not affect) the resulting SkX is an interesting open question. The SkX of Ref.\cite{ozawa2017zero} can be stabilized even in zero filed, while the present SkX requires finite magnetic fields for stabilization.

 In addition to the presence/absence of the RSB discussed above, if one compares the properties of the long-range RKKY model with those of the 2D short-range \cite{okubo2012multiple}, there are some differences in spite of the basic similarity in their phase structures.

 First, the 2D short-range model exhibits the $Z$ phase, the random domain state consisting of the SkX and the anti-SkX, while there is no indication of the $Z$ phase in our present computation, at least in the range of sizes studied. One possibility might be a finite-size effect. As the stabilization of the $Z$ phase requires minimum lattice sizes to accommodate several SkX and anti-SkX domains, typically of the linear size 10$\sim$20 lattice spacings, our present maximum lattice size might not be sufficiently large to stabilize the $Z$ phase. The other possibility might be that the absence of the $Z$ phase is a real effect. The $Z$ phase is stabilized by fluctuations, not describable by the Landau mean-filed theory. Generally speaking, since the high spatial dimensionality and the long-range interaction tend to suppress fluctuations, the $Z$ phase might well be absent even in the bulk limit in the present 3D RKKY model.

 In the $T$-$h$ phase diagram of the 2D short-range model, the single-$q$ phase extends toward high temperature forming a narrow band between the double-$q$ phase and the paramagnetic (field-induced magnetized) phase. Such a feature is absent in the present 3D RKKY model. Presumably, this is due to the occurrence of the RSB in the RSB double-$q$ phase where the single-$q$ state already takes considerable weight macroscopically coexisting with the double-$q$ state, in contrast to the double-$q$ state of the 2D short-range model which is replica-symmetric and cannot contain the single-$q$ state with a nonzero weight.

 To the authors' knowledge, the present model is the first example exhibiting the RSB in the regularly ordered states with spatial periodicity. The present RSB realized in the regularly ordered phase has a unique simplicity not shared by the standard RSB in glassy ordered states, e.g., spin glasses and molecular or structural glasses. One can specify each constituent pure state by the standard order parameter, e.g., the chiralities and the $Z_3$-symmetry-breaking parameter, in contrast to the standard RSB in glassy systems where each constituent pure state is rather complex without any periodicity and is difficult to be specified. Especially in the case of the triple-$q$ SkX phase, the number of constituent pure states is finite when viewed from the discrete symmetries, in contrast to an infinite number of pure states in glassy systems. 

 Finally, we wish to discuss the possible experimental observation of the RSB feature. How the RSB features manifest themselves in experiments is an important but subtle question. First, let us consider an ideal situation without any perturbative interaction nor impurities or imperfections. In the RSB SkX phase of such an ideal system, either the SkX state or the single-$q$ state is realized by chance. Experimentally, this would mean that, depending on the initial conditions of the measurements and the details of each run, either the triple-$q$ SkX state or the single-$q$ state is realized by chance, each with a finite probability. If we could perform cooling experiments in such an ideal situation many times from the paramagnetic phase to the RSB SkX phase by using the same sample and the same protocol, and measure the topological Hall effect in each run, we would obtain the characteristic distribution function of the anomalous Hall conductivity, with a central peak corresponding to the single-$q$ spiral state and the symmetric positive and negative peaks corresponding to the SkX state and the anti-SkX state.

 Of course, the real situation could be more complex. In reality, the sample contains weak perturbative interactions, e.g., the dipolar interaction, which are likely to discriminate the constituent states of the RSB. If the energy difference between the constituent states is larger than the thermal energy $\sim k_{\rm B}T$, only the preferable state could be realized in reality. If the energy difference is smaller than or comparable to the thermal energy $\sim k_{\rm B}T$, all the constituent states of RSB would be realized, perhaps with a certain bias in its realization probability. In the presence of impurities and imperfections, which is inevitable in real samples, these constituent states would form macroscopic or semi-macroscopic domains, which might be more or less pinned by these impurities and imperfections. Hence, in reality, the RSB would appear in the form of macroscopic (or semi-macroscopic) domains consisting of both the single-$q$ and the triple-$q$ SkX (or the double-$q$) states. How to detect and control these macroscopic (semi-macroscopic) domains would then be an interesting and challenging future problem.

\begin{acknowledgments}
The authors would like to thank K. Aoyama, R. Osamura and J. Takahashi for useful discussion. 
We are thankful to ISSP, the University of Tokyo, and YITP, Kyoto University for providing us with CPU time. 
This work is supported by JSPS KAKENHI Grant No. JP17H06137.
\end{acknowledgments}

\begin{widetext}

\appendix

\section{Ewald sum of the RKKY interaction on a stacked triangular lattice}
\label{sec_ewald_app}

 In this subsection, we give some of the details of the application of the Ewald-sum method to our Hamiltonian Eq. (\ref{hamiltonian}). In this method, we take account of the long-range RKKY interaction beyond the finite-system size $L$ in the form adapted to the imposed periodic boundary conditions,
\begin{eqnarray}
J_{ij}^{\rm Ewald} &=& -J_0 a^3 \sum_{\bm{\lambda}} J_{ij}(\bm{ \lambda}), \label{ewald} \\
J_{ij}(\bm{\lambda}) &=& \frac{\cos (2k_{\rm F} |\bm{r}_{ij} + L \bm{\lambda}|)}{|\bm{r}_{ij} + L \bm{\lambda}|^3} - \frac{\sin (2k_{\rm F} |\bm{r}_{ij} + L \bm{\lambda}|)}{2k_{\rm F}|\bm{r}_{ij} + L \bm{\lambda}|^4},
\end{eqnarray}
where $L$ is linear size of the system, and
\begin{equation}
\bm{\lambda} = n_a \bm{a} + n_b \bm{b} + \frac{L_z n_c}{L} \bm{c},
\end{equation}
with $\bm{a} = (1,0,0),~\bm{b} = (1/2,\sqrt{3}/2,0),~\bm{c} = (0,0,c)$. The sum in Eq. (\ref{ewald}) runs over integers $n_\mu = -\infty,...,0,...\infty$ ($\mu = a,b,c)$, $L \bm{\lambda}$ mapping the original cell of the size $L\times L\times L_z$ to the image cell with exactly the same spin configuration as that in the original cell.

Noting the identity
\begin{equation}
1 = \frac{1}{\Gamma (\frac{3}{2})} \left[ \Gamma \left( \frac{3}{2},\pi\frac{|\bm{r}_{ij} + L \bm{\lambda}|^2}{L^2} \right) + \gamma \left( \frac{3}{2},\pi\frac{|\bm{r}_{ij} + L \bm{\lambda}|^2}{L^2}\right) \right],
\label{identity}
\end{equation}
where $\Gamma(\alpha) = \int_0^\infty dt e^{-t} t^{\alpha -1}$ is the gamma function, $\Gamma(\alpha, x) = \int_x^\infty dt e^{-t} t^{\alpha -1}$ and $\gamma(\alpha, x) = \int_0^x dt e^{-t} t^{\alpha -1}$ are the incomplete gamma functions. $\Gamma(\alpha, x)$ and $\gamma(\alpha, x)$ take large values for small $x$ and for large $x$, respectively. We insert Eq. (\ref{identity}) to Eq. (\ref{ewald}) and obtain,
\begin{eqnarray}
J_{ij}^{\rm Ewald} &=& -\frac{J_0 a^3}{\Gamma(3/2)} (C_{ij}^{\rm short} + C_{ij}^{\rm long} + S_{ij}^{\rm short} + S_{ij}^{\rm long}), \label{j_ewald} \\
C_{ij}^{\rm short} &=& \sum_{\bm{\lambda}} \Gamma \left(\frac{3}{2},\pi\frac{|\bm{r}_{ij} + L \bm{\lambda}|^2}{L^2} \right) \frac{\cos (2k_{\rm F} |\bm{r}_{ij} + L \bm{\lambda}|)}{|\bm{r}_{ij} + L \bm{\lambda}|^3}, \\
C_{ij}^{\rm long} &=& \sum_{\bm{\lambda}} \gamma \left(\frac{3}{2},\pi\frac{|\bm{r}_{ij} + L \bm{\lambda}|^2}{L^2} \right) \frac{\cos (2k_{\rm F} |\bm{r}_{ij} + L \bm{\lambda}|)}{|\bm{r}_{ij} + L \bm{\lambda}|^3},\\
S_{ij}^{\rm short} &=& -\sum_{\bm{\lambda}} \Gamma \left(\frac{3}{2},\pi\frac{|\bm{r}_{ij} + L \bm{\lambda}|^2}{L^2} \right) \frac{\sin (2k_{\rm F} |\bm{r}_{ij} + L \bm{\lambda}|)}{2k_{\rm F}|\bm{r}_{ij} + L \bm{\lambda}|^4},\\
S_{ij}^{\rm long} &=& -\sum_{\bm{\lambda}} \gamma \left(\frac{3}{2},\pi\frac{|\bm{r}_{ij} + L \bm{\lambda}|^2}{L^2} \right) \frac{\sin (2k_{\rm F} |\bm{r}_{ij} + L \bm{\lambda}|)}{2k_{\rm F}|\bm{r}_{ij} + L \bm{\lambda}|^4}.
\end{eqnarray}
$C_{ij}^{\rm short}$ and $S_{ij}^{\rm short}$ are well-converging functions with respect to $|\bm{r}_{ij} + L \bm{\lambda}|$. To make $C_{ij}^{\rm long}$ and $S_{ij}^{\rm long}$ converge faster, we deal with them in the Fourier space. For example, $C_{ij}^{\rm long}$ can be rewritten as
\begin{eqnarray}
C_{ij}^{\rm long} &=& \int d^3 \bm{\rho} \sum_{\bm{\lambda}} \delta(\bm{\rho}-L\bm{\lambda}) \gamma \left(\frac{3}{2},\pi\frac{|\bm{r}_{ij} + \bm{\rho}|^2}{L^2} \right) \frac{\cos (2k_{\rm F} |\bm{r}_{ij} + \bm{\rho}|)}{|\bm{r}_{ij} + \bm{\rho}|^3} \nonumber \\
&=& \frac{1}{\mathcal{J}} \int d^3 \bm{k} \sum_{\bm{h}} \delta \left(\bm{k} - \frac{\bm{h}}{L} \right) \int  d^3 \bm{\rho} e^{-2\pi i \bm{k}\cdot \bm{\rho}} 
\gamma \left(\frac{3}{2},\pi\frac{|\bm{r}_{ij} + \bm{\rho}|^2}{L^2} \right) \frac{\cos (2k_{\rm F} |\bm{r}_{ij} + \bm{\rho}|)}{|\bm{r}_{ij} + \bm{\rho}|^3} \nonumber \\
&=& \frac{1}{\mathcal{J}} \int d^3 \bm{k} \sum_{\bm{h}} \delta \left(\bm{k} - \frac{\bm{h}}{L} \right) \frac{e^{2\pi i \bm{k} \cdot \bm{r}_{ij}}}{k} 
\int_0^\infty d\rho \frac{\gamma \left(\frac{3}{2},\pi\frac{\rho^2}{L^2} \right)}{\rho^2} \left[\sin (2\pi k_+ \rho) + \sin (2\pi k_- \rho) \right],
\end{eqnarray}
where
\begin{equation}
\bm{h} = m_a \bm{k}_a + m_a \bm{k}_a + \frac{L}{L_z} m_c \bm{k}_c ,
\end{equation}
and $k_\pm = k \pm k_{\rm F}/\pi$. The sum $\sum_{\bm{h}}$ runs over all integers $m_\mu = -\infty,...,0,...\infty$ ($\mu = a,b,c$). In the deformation from the fist line to the second line, we used the Parseval's theorem,
\begin{equation}
\int d^3 \bm{\rho} A (\bm{\rho}) B (\bm{\rho}) = \int d^3 \bm{k} \tilde{A} (\bm{k}) \tilde{B} (\bm{k}),
\end{equation}
where $\tilde{f} (\bm{k}) = \int d^3 \bm{\rho} f(\bm{\rho}) e^{-2\pi i \bm{k} \cdot \bm{\rho}}$ is the Fourier transform of $f(\bm{\rho})$, and the Poisson's summation formula, 
\begin{equation}
\sum_{\bm{\lambda}} e^{2\pi i \bm{k}\cdot L\bm{\lambda}} = \sum_{m_a,m_b,m_c}\delta (L \bm{k} \cdot \bm{a} - m_a)\delta (L \bm{k} \cdot \bm{b} - m_b)\delta (L_z \bm{k} \cdot \bm{c} - m_c) = \frac{1}{\mathcal{J}} \sum_{\bm{h}}\delta \left(\bm{k} - \frac{\bm{h}}{L} \right),
\end{equation}
where $\mathcal{J} =  \frac{\sqrt{3}}{2}L \times L \times L_z$ is the Jacobian determinant. Applying a formula \cite{nijboer1957calculation}, 
\begin{equation}
\int_0^\infty d \rho \frac{\gamma (\frac{3}{2}, \pi \frac{\rho^2}{L^2})}{\rho^2}\sin (2\pi k \rho) = \frac{\pi^{\frac{3}{2}}k}{2} E_1 (\pi k^2 L^2),
\end{equation}
where $E_1(x) = \int_x^\infty dt e^{-t}/t$ is the exponential integral function, we get $C_{ij}^{\rm long}$ as
\begin{equation}
C_{ij}^{\rm long} = \frac{1}{\mathcal{J}} \sum_{\bm{h}} \frac{\pi^{\frac{3}{2}}e^{2\pi i \frac{\bm{h}}{L}\cdot \bm{r}_{ij}}}{2 h} \left[h_+ E_1 (\pi h_+^2) - h_- E_1 (\pi h_-^2) \right],
\end{equation}
where $h=|h|$ and $h_\pm = h \pm L k_{\rm F}/\pi$. Since $E_1(x)$ is a well-converging function, we can take the sum $\sum_{\bm{h}}$ numerically with high precision. 
 
 In the same manner, $S_{ij}^{{\rm long}}$ can be written as
\begin{equation}
S_{ij}^{\rm long} = -\frac{1}{2k_{\rm F}\mathcal{J}} \sum_{\bm{h}} \frac{\pi^{\frac{3}{2}}e^{2\pi i \frac{\bm{h}}{L}\cdot \bm{r}_{ij}}}{2 h L} \left[ \pi h_+^2  E_1 (\pi h_+^2) -   h_-^2 E_1 (\pi h_-^2) - (e^{-\pi h_+^2} - e^{-\pi h_-^2}) \right].
\end{equation}
\begin{figure}[t]
\includegraphics[clip,width=120mm]{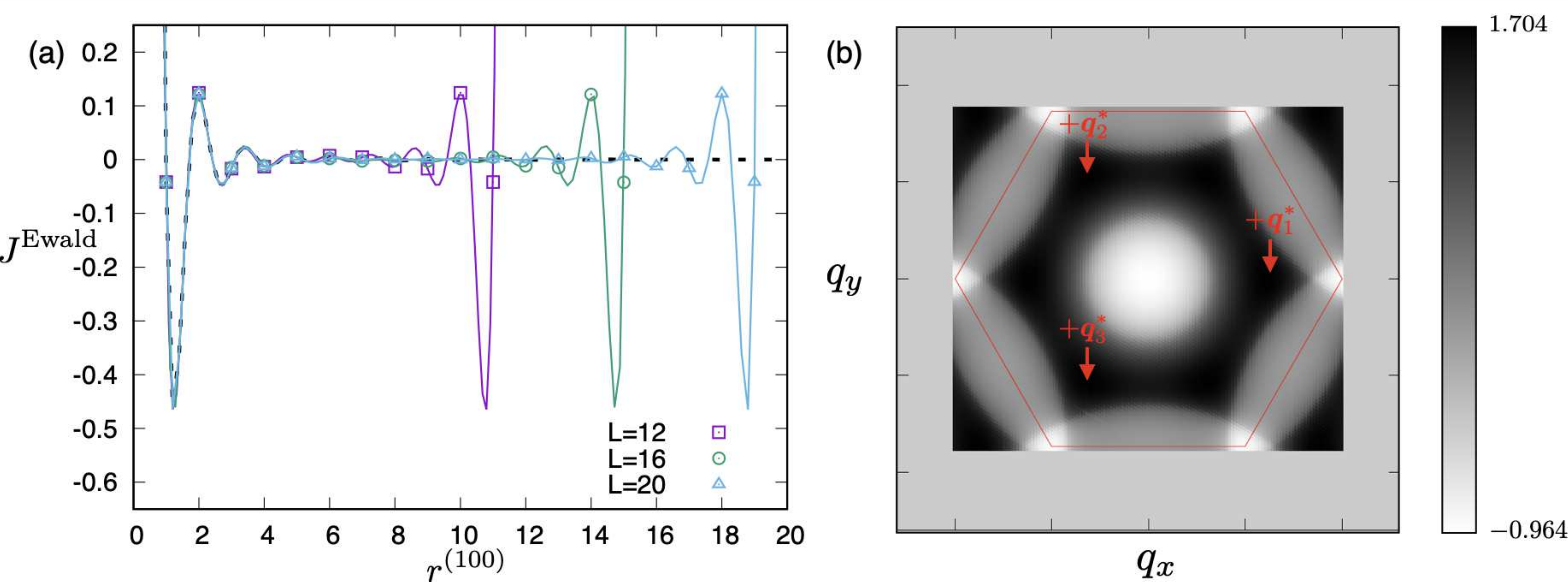}
\caption{(a): The RKKY Ewald potential is plotted versus the spin distance along the $(100)$ direction for the incommensurate case of $k_{\rm F} = 2\pi/2.77$ and $c=1.45$. The dashed line is the bare RKKY interaction without the Ewald sum. (b): The 2D intensity map of the Fourier transform of the RKKY Ewald potential in the $(q_x,q_y)$ plane with $q_z=0$ for the incommensurate case of  $k_{\rm F} = 2\pi/2.77$ and $c=1.45$. The lattice size $L=96$. The red hexagon represents the first Brillouin zone.
}
\label{ewald_app}
\end{figure}

 For the parameter choice of our present MC simulation on the RKKY Hamiltonian, $k_F=2\pi/2.77$ and $c=1.45$, we show in Fig. \ref{ewald_app} (a) the Ewald periodic potential $J_{ij}^{\rm Ewald}$ in units of $J_0a^3$ as a function of the spin distance along the $(100)$ direction. The potential has a symmetric form with respect to $r^{(100)} = L/2$ due to the applied periodic boundary conditions. In Fig.\ref{ewald_app} (b), we show the Fourier transform of $J_{ij}^{\rm Ewald}$, defined by
\begin{equation}
J_{\bm{q}}^{\rm Ewald}(\bm{r}_{ij}) = \frac{1}{N} \sum_{j=1}^N J_{ij}^{\rm Ewald}(\bm{q}) e^{-i \bm{q} \cdot\bm{r}_{ij}} ,
\end{equation}
in the $(q_x, q_y)$-plane with $q_z=0$. For our present choice of the parameters, $k_F=2\pi/2.77$ and $c=1.45$, the maximum intensities are located at the incommensurate wavenumbers, $\pm \bm{q}_1^*, \pm \bm{q}_2^*, \pm \bm{q}_3^*$, as indicated by the arrows in Fig. \ref{ewald_app} (b).

\section{The temperature and magnetic-field dependence of physical quantities}
\label{sec_mag_temp_app}

\begin{figure}[t]
\includegraphics[clip,width=140mm]{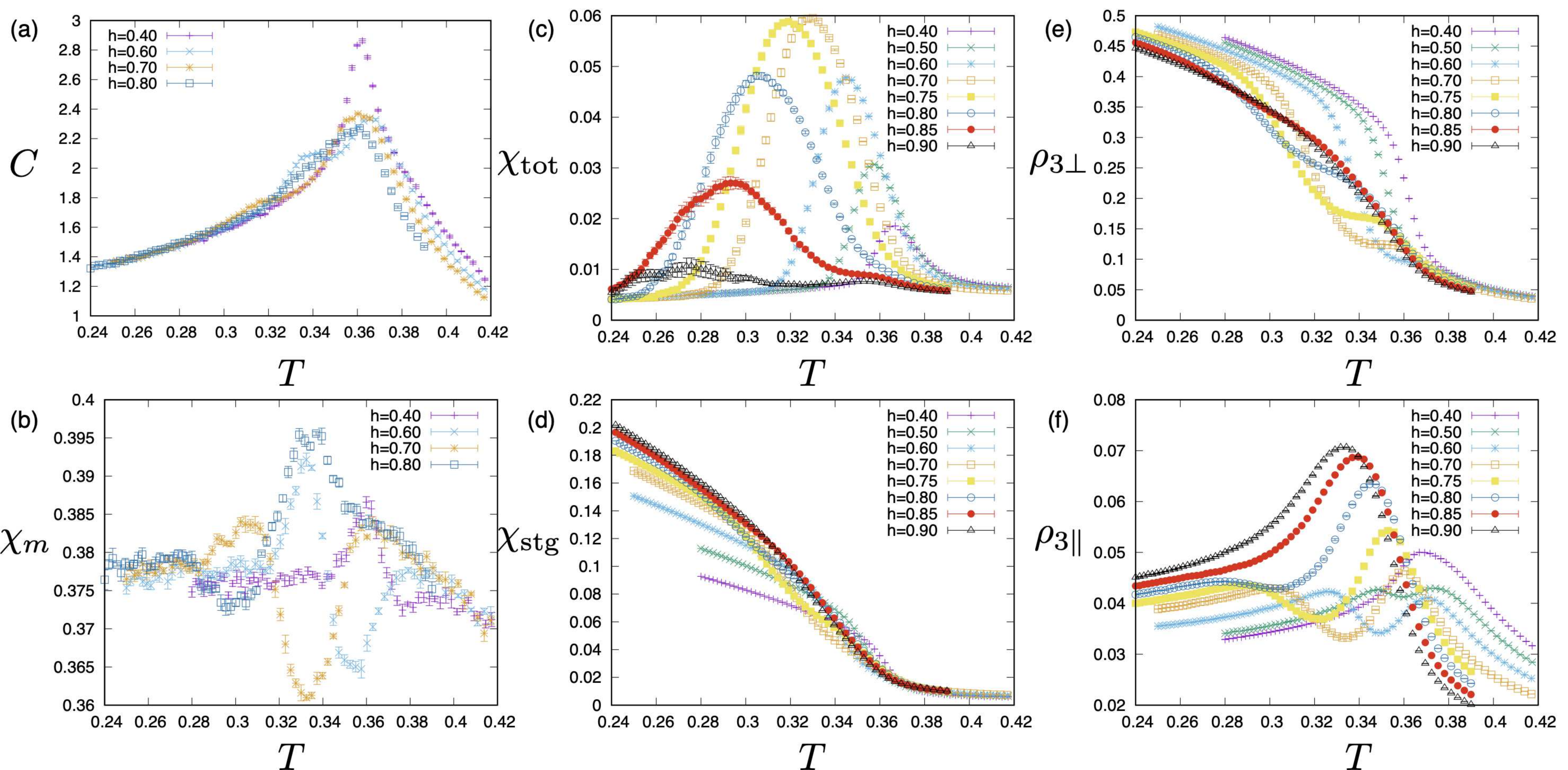}
\caption{The temperature and magnetic-field dependence of physical quantities. The lattice size is $L=20$. (a) The specific heat, (b) the magnetic susceptibility along the field direction, (c) the total scalar chirality, (d) the staggered scalar chirality, (e) the $Z_3$-symmetry-breaking parameter for the perpendicular spin component, and (f) the $Z_3$-symmetry-breaking parameter for the parallel spin component.
}
\label{field_app}
\end{figure}

 In this subsection, we present the temperature and magnetic-field dependence of several physical quantities which supplements the data shown in the main text. Figs. \ref{field_app} (a)-(f) exhibit the temperature dependence of various physical quantities under magnetic fields, including the magnetic susceptibility along the magnetic-field direction ($S_z$-direction) calculated from the magnetization fluctuation. As can be seen from Fig. \ref{field_app} (b), the magnetic susceptibility in the RSB SkX phase tends to be suppressed, exhibiting a dip feature there.

\section{Spin and chirality configurations in real space in the double-$q$ state} 
\label{sec_double_spin_chiral}

\begin{figure}[t]
\includegraphics[clip,width=140mm]{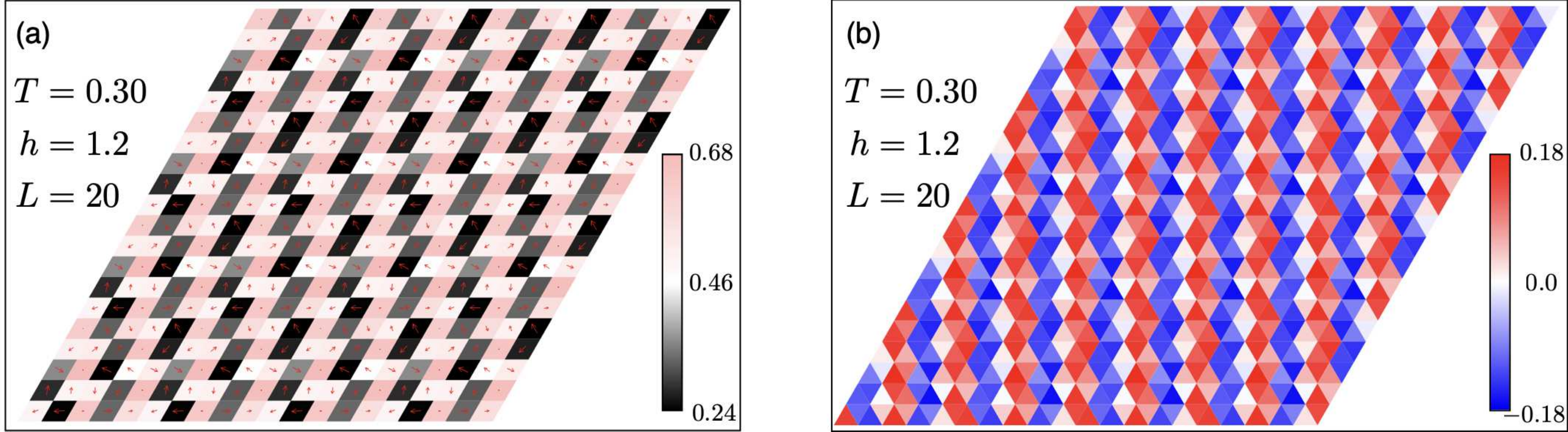}
\caption{Typical real-space (a) spin and (b) chirality configurations of the double-$q$ state, taken from MC simulations of the RSB double-$q$ phase at ($T=0.30$, $h=1.2$). To reduce the thermal noise, the short-time averaging over 100 MCS is made by using the Metropolis updating only.
}
\label{double_app2}
\end{figure}

 In this subsection, we show the spin and the chirality configurations in real space for the double-$q$ state. In Fig. \ref{double_app2}, for a typical double-$q$ state realized as an equilibrium state in the RSB double-$q$ phase at ($T=0.30$, $h=1.2$), we show the real-space (a) spin and (b) chirality configurations. The associated spin structure factors are given in Figs. \ref{snap_double3} (c) and (d) of the main text, each for the perpendicular and the parallel components. As can be seen from Fig. \ref{double_app2}, both the spin $S_z$-component and the scalar chirality form a linear spin-density-wave pattern along the $x$-direction, just corresponding to the single-$q$-like pattern of $S_\parallel(\bm{q})$ shown in Fig. \ref{snap_double3} (d). As can be seen from such chirality configuration shown in Figs. \ref{double_app2} (b), both the total and the staggered scalar chiralities $\chi_{{\rm tot}}$ and $\chi_{{\rm stg}}$ vanish in the double-$q$ state.

\section{Mean-field calculation for the RSB double-$q$ phase}
\label{sec_mean_app}

 In this subsection, we present the results of our mean-field calculation performed to better understand the character of the RSB double-$q$ phase, with particular interest in the intermediate states apparently connecting the pure single-$q$ and the pure double-$q$ states.

 The Landau free energy up to the forth order is given by
\begin{eqnarray}
\frac{F}{N} = \frac{1}{2}\sum_{\bm{q}} [3T-J_{\bm{q}}]|\bm{B}_{\bm{q}}|^2 - H B_{0,z} + \frac{9T}{20}{\sum_{\bm{q}_1\sim \bm{q}_4}}^\prime[\bm{B}_{\bm{q}_1}\cdot\bm{B}_{\bm{q}_2}][\bm{B}_{\bm{q}_3}\cdot\bm{B}_{\bm{q}_4}] ,
\label{landau}
\end{eqnarray}
where
\begin{equation}
\bm{B}_{\bm{q}} = \frac{1}{N}\sum_{\bm{r}}\bm{B}(\bm{r})\exp(-i\bm{q}\cdot \bm{r})
\end{equation}
is the Fourier component of the spin field $\bm{B}(\bm{r}) = \langle \bm{S}(\bm{r}) \rangle$ \cite{reimers1991mean, okubo2012multiple}. The sum ${\sum_{\bm{q}_1\sim \bm{q}_4}}^\prime$ runs over $\bm{q}_1\sim \bm{q}_4$'s satisfying the constraint $\bm{q}_1 + \bm{q}_2 + \bm{q}_3 + \bm{q}_4 = \bm{0}$. The mea-field transition temperature is given by $T_{\rm c} = \frac{1}{3}J_{\bm{q}^*}$. Below $T_c$, we consider only the six incommensurate modes $\bm{q} = \pm \bm{q}_1^*,\pm \bm{q}_2^*,\pm \bm{q}_3^*$ and the uniform mode $\bm{q} = \bm{0}$, where $\bm{q}_1^*$, $\bm{q}_2^*$ and $\bm{q}_3^*$ are the $\bm{q}$-values giving the $J(\bm{q})$ maxima ($\bm{q}_1^* + \bm{q}_2^* + \bm{q}_3^* = \bm{0}$). 

 Now, we restrict the phase space to that relevant to the RSB double-$q$ phase, and assume
\begin{eqnarray}
&\bm{B}_{0}& \parallel \bm{e}_z,\ \  \bm{B}_{\pm 1} \parallel \bm{e}_z, \\
\bm{B}_{\pm 2} \perp &\bm{e}_z&,\ \bm{B}_{\pm 3} \perp \bm{e}_z,\ \ \ \  \bm{B}_{\pm 2} \perp \bm{B}_{\pm 3}.
\end{eqnarray}
 Putting $|\bm{B}_0| = m_0,~ |\bm{B}_{\pm 1}| = m_1,~ |\bm{B}_{\pm 2}| = m_2,~ |\bm{B}_{\pm 3}| = m_3$, and using the abbreviation $\bm{B}_{\pm \bm{q}_i}=\bm{B}_{\pm i}$ ($i=1,2,3$), the quartic term of the free energy in Eq. (\ref{landau}) can be written as,
\begin{eqnarray}
\nonumber
f_4
&=& m_0^4 + 4m_0^2m^2 + 4m^4 + 8m_0^2m_1^2  \nonumber \\
&+& 2\sum_i |\bm{B}_{+i} \cdot \bm{B}_{+i}|^2 + 8\sum_{i \neq j} \left[ |\bm{B}_{+i} \cdot \bm{B}_{+j}|^2 + |\bm{B}_{+i} \cdot \bm{B}_{-j}|^2 \right] \nonumber \\
&+& 8\left[ [\bm{B}_{0}\cdot \bm{B}_{+1}][\bm{B}_{+2} \cdot \bm{B}_{+3}] + [\bm{B}_{0}\cdot \bm{B}_{-1}][\bm{B}_{-2} \cdot \bm{B}_{-3}] \right] ,
\end{eqnarray}
where
\begin{equation}
m^2 = m_1^2 + m_2^2 + m_3^2.
\label{constraint}
\end{equation}
The quartic term of the free energy $f_4$ is minimized when $|\bm{B}_{\pm 2} \cdot \bm{B}_{\pm 2}| = |\bm{B}_{\pm 3} \cdot \bm{B}_{\pm 3}| = |\bm{B}_{\pm 2} \cdot \bm{B}_{\mp 3}| = 0$. Explicit forms of $\bm{B}_{\pm 1}$, $\bm{B}_{\pm 2}$ and $\bm{B}_{\pm 3}$ satisfying these conditions are given by 
\begin{eqnarray}
\bm{B}_{\pm 1} &=& (0,0,m_1 e^{\pm i\theta_1}) , \\
\bm{B}_{\pm 2} &=& \frac{m_2}{\sqrt{2}}(e^{\pm i \theta_2},  e^{\pm i (\theta_2 - \frac{\pi}{2})}, 0) , \\
\bm{B}_{\pm 3} &=& \frac{m_3}{\sqrt{2}}(e^{\pm i \theta_3},  e^{\pm i (\theta_3 + \frac{\pi}{2})}, 0) ,
\end{eqnarray}
where $\theta_1$, $\theta_2$ and $\theta_3$ are phase factors of the modes 1, 2 and 3, respectively. In the real space, they are given by
\begin{equation}
\bm{B}(\bm{r}) = \left(
\begin{array}{c}
\sqrt{2}m_2\cos(\bm{q}_1^*\cdot \bm{r} + \theta_2) + \sqrt{2}m_3\cos(\bm{q}_3^*\cdot \bm{r} + \theta_3) \\
\mp \sqrt{2}m_2\sin(\bm{q}_2^*\cdot \bm{r} + \theta_2) \pm \sqrt{2}m_3\sin(\bm{q}_3^*\cdot \bm{r} + \theta_3) \\
m_0 + 2m_1\cos(\bm{q}_1^*\cdot \bm{r} + \theta_1)
\end{array}
\right).
\label{single_double}
\end{equation}
Eq. (\ref{single_double}) represents the single-$q$ state if $m_1=m_2 = 0$, while it represents the double-$q$ state if $m_1 > 0$ and $m_2=m_3 > 0$. If $m_1>0$ and $m_2\neq m_3>0$, it represents the intermediate state. For these $\bm{B}_{\pm i}$'s, $f_4$ is given by
\begin{eqnarray}
f_4 =  m_0^4 + 4m_0^2m^2 + 4m^4 + 8m_0^2m_1^2 + 2m_1^4 +8m_2^2m_3^2 + 16m_0m_1m_2m_3\cos (\theta_1+\theta_2+\theta_3),
\label{f_4double}
\end{eqnarray}
which is minimized for 
\begin{equation}
\cos(\theta_1 + \theta_2 + \theta_3) = -1,
\end{equation}
yielding 
\begin{eqnarray}
f_4 =  m_0^4 + 4m_0^2m^2 + 4m^4 + 8m_0^2m_1^2 + 2m_1^4 +8m_2^2m_3^2 - 16m_0m_1m_2m_3.
\label{f_4double2}
\end{eqnarray}

 Once $m$ and $m_0$ are given, which can be regarded as the measure of the temperature and the applied magnetic field,  $f_4$ under the constraint Eq. (\ref{constraint}) becomes only the function of the ``mixing ratio'' $u$, i.e., the ratio between the amplitudes of the two in-plane modes $m_2$ and $m_3$, 
\begin{equation}
u = \frac{m_2}{m_3}\ \ \ (0 \leq u \leq 1) ,
\end{equation}
where $u=0$ and $u=1$ represent the single-$q$ and the double-$q$ states, respectively.

\begin{figure}[t]
\includegraphics[clip,width=120mm]{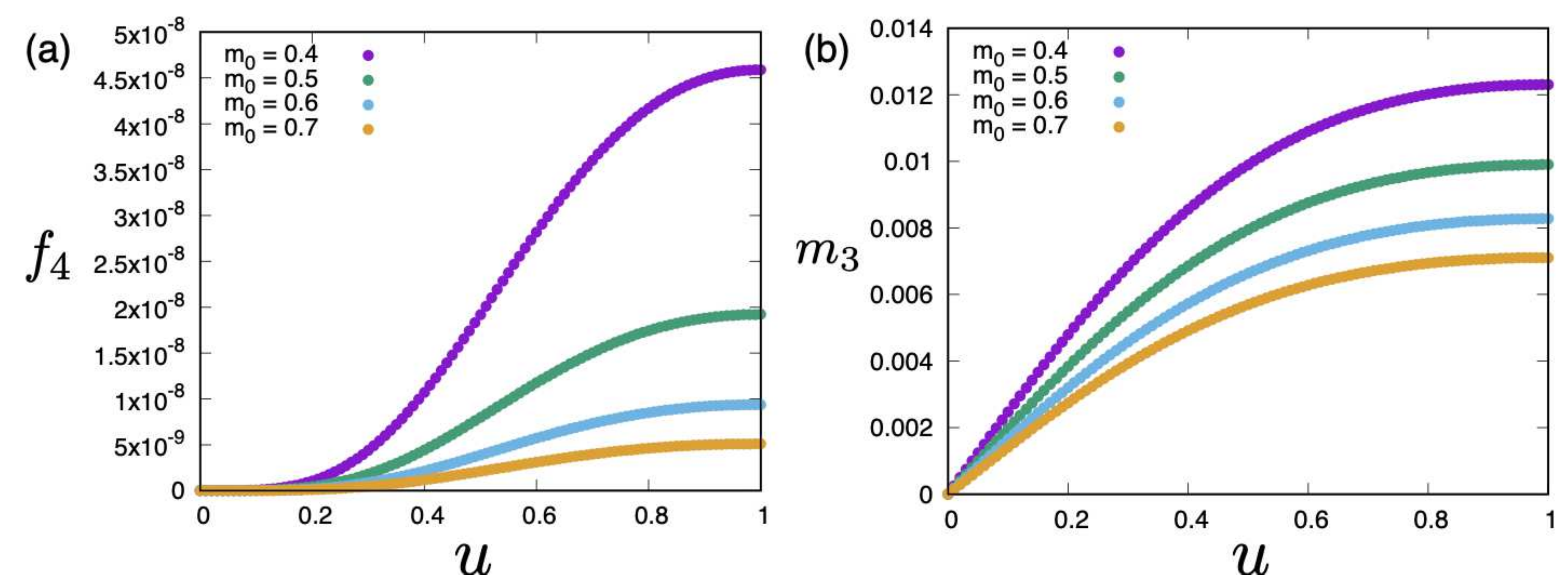}
\caption{The mixing-rate parameter $u$-dependence of (a) the quartic part of the free energy, $f_4$, and of (b) the amplitude of the $\bm{q}_3$-mode associated with the linear density wave, $m_3$, where $u=0$ corresponds to the pure single-$q$ state, $u=1$ to the pure double-$q$ state, and $0<u<1$ to the intermediate state.
}
\label{fene_app}
\end{figure}
\begin{figure}[t]
\includegraphics[clip,width=120mm]{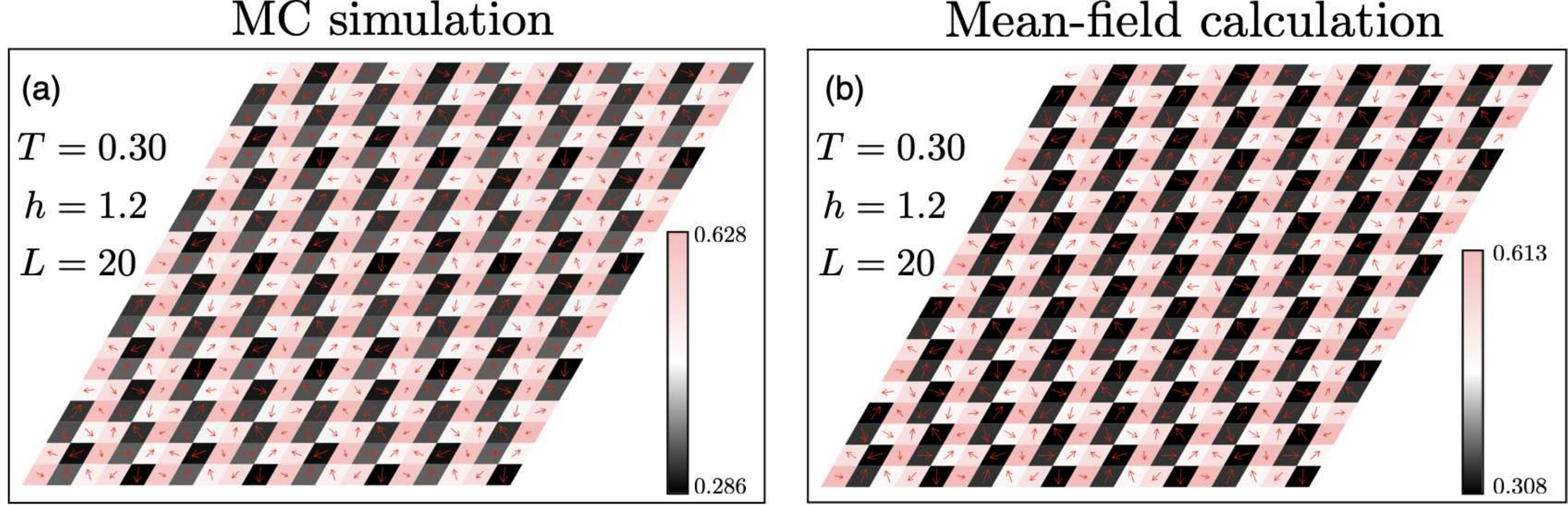}
\caption{Real-space spin configurations prepared (a) from MC simulations by Fourier-converting the $S(\bm{q})$ intensities at $\bm{q} = \bm{q}_1^*, \bm{q}_2^*, \bm{q}_3^*$ shown in Figs. \ref{snap_double3} (e) and (f), and (b) from the mean-field formula Eq. (\ref{single_double}) by properly determining its parameters: See the text of Appendix \ref{sec_mean_app} for details). 
}
\label{snap_double_app}
\end{figure}

 In Fig. \ref{fene_app} (a), we show the mixing-ratio $u$-dependence of $f_4$ computed for $m=0.1$ for several values of $m_0$. It takes a minimum for the single-$q$ state with $u=0$, takes a maximum for the double-$q$ state with $u=1$, and connects these two ends via the intermediate states with $0 < u < 1$. At the mean-field level, the single-$q$ state has been known to give a global minimum \cite{okubo2012multiple}, and the same situation arises in our present result. Fig. \ref{fene_app} (b) exhibits the $u$-dependence of $m_3$, which becomes zero for the single-$q$ state and becomes nonzero when the double-$q$ state is mixed with a nonzero portion. 

 Although the mean-field analysis cannot provide an ordered state corresponding to the true free-energy minimum because of its inadequacy to take account of the fluctuation effect, it still gives useful information for the intermediate states we have found in the RSB double-$q$ state by MC simulations. Thus, we compare the spin configuration of the intermediate state realized as an equilibrium state of the RSB double-$q$ phase in our MC simulation with those of the mean-field calculation given by Eq. (\ref{single_double}). 

 In Fig. \ref{snap_double_app} (a), we show the real-space spin configuration from our MC simulation corresponding to the spin structure factor $S(\bm{q})$ shown in Figs. \ref{snap_double3} (e) and (f), which is prepared by Fourier-converting the observed $S(\bm{q})$ intensities at $\bm{q} = \bm{q}_1^*, \bm{q}_2^*, \bm{q}_3^*$. For comparison, we show in Fig. \ref{snap_double_app} (b) the real-space spin configuration obtained from the mean-field calculation, Eq. (\ref{single_double}), where the coefficients $m_0$, $m_1$, $m_2$ and $m_3$ are determined from $S(\bm{q})$ of Figs. \ref{snap_double3} (e) and (f) as $m_0 = S_{\parallel}(\bm{0}),~m_1 = S_{\parallel}(\bm{q}_1),~m_2 = S_{\perp}(\bm{q}_2),~m_3 = S_{\parallel}(\bm{q}_3)$. As can be seen from the figure, the two spin configurations (a) and (b) resemble quite well, indicating that Eq. (\ref{single_double}) well describes the spin configuration of the intermediate state in the RSB double-$q$ phase.

\end{widetext}

\bibliography{rkky_ref}

\end{document}